\documentclass[aps, physrev, reprint, superscriptaddress,noeprint,hypens,floatfix]{revtex4-1}

\usepackage[bookmarks=true,colorlinks=true,urlcolor=blue,linkcolor=blue,citecolor=blue,breaklinks=true]{hyperref}
\PassOptionsToPackage{hyphens}{url}
\usepackage{graphicx}
\usepackage[utf8]{inputenc}
\usepackage{amsmath, amssymb}
\usepackage[table]{xcolor}       
\newcommand{\DIPC}[0]{
Donostia International Physics Center (DIPC),
Paseo Manuel de Lardizabal 4, 20018 Donostia-San Sebasti\'an, Spain}
\newcommand{\CFM}[0]{
Centro de F\'{\i}sica de Materiales CFM/MPC (CSIC-UPV/EHU), Paseo Manuel de Lardizabal 5, 20018 Donostia-San Sebasti\'an, Spain}
\newcommand{\PolymerEHU}[0]{Departamento de Pol\'{i}meros y Materiales Avanzados: F\'{i}sica, Qu\'{i}mica y Tecnolog\'{i}a, Facultad de Qu\'{i}micas (UPV/EHU), Apartado 1072, 20080 Donostia-San Sebasti\'{a}n, Spain}

\begin{document}

\title{Adsorption and dissociation of diatomic molecules in monolayer $1H$-MoSe$_2$}

\author{Raúl Bombín}
\email{raul.bombin@ehu.eus}
\affiliation{\PolymerEHU}
\affiliation{\CFM}

\author{Maite Alducin}
\email{maite.alducin@ehu.eus}
\affiliation{\CFM}
\affiliation{\DIPC}

\author{J. I\~{n}aki Juaristi}
\email{josebainaki.juaristi@ehu.eus}
\affiliation{\PolymerEHU}
\affiliation{\CFM}
\affiliation{\DIPC}

\date{\today}

\begin{abstract}
Two dimensional transition metal dichalcogenides appear as good candidates for gas sensing and catalysis. Here, by means of density functional theory, we characterize the adsorption and dissociation of selected diatomic molecules (CO, H$_2$, O$_2$, and NO) on the MoSe$_2$ monolayer. We consider that these processes occur on the pristine $1H$-MoSe$_2$ monolayer and in the vicinity of an isolated Se vacancy. The presence of a single Se vacancy both enhances the molecular adsorption and reduces the energy needed for dissociation, making it energetically favorable for the case of  O$_2$ and NO molecules. Moreover, the presence of a second Se vacancy makes the dissociation process energetically favorable for all the molecules that are studied here. 
For each case we evaluate the effect that each adsorbate has on the electronic structure of the MoSe$_2$ monolayer and the charge transfer that takes place between the adsorbate and the surface. Remarkably, adsorption of either CO or NO at the Se vacancy induces a finite spin-magnetization in the system that is spatially well localized around the adsorbate and the three closest Mo atoms.
\end{abstract}

\maketitle

\section{Introduction}

Atomically thin nano-structures have attracted much attention both from academia and industry since the discovery of graphene in 2004~\cite{Novoselov666}.  However, the zero band-gap that graphene exhibits has emerged as its main drawback for practical applications~\cite{Guanxiong2013} due to the large leakage currents. 
Nonetheless, the advantages that two-dimensional (2D) materials offer, such as large surface to volume ratio and good implementation with metal electrodes~\cite{Tyagi2020}, has boosted the research in other families of 2D materials.

Transition Metal Dichalcogenides (TMDs) appear in nature in a bulk, layered structure. Each layer is  bounded to the others by weak van der Waals interactions and it is composed of planes of transition metal atoms (M) sandwiched between two layers of chalcogen ones (X$=$S, Se, and Te). This allows to exfoliate them into single and few layered nano-structures. In addition to the different chemical compositions in which they appear (MX$_2$), TMDs present different phases depending on the coordination between the metal and the chalcogen atoms (trigonal prismatic, octahedral, and distorted octahedral). Besides that, they are very versatile as their electrical properties can be tuned by, for example, application of strain \cite{Muoi2019}, application of an external electric field~\cite{Ai2019}, and creation of vacancies~\cite{Santos2020,Liang2021}.

Group V TMDs (i.e., MX$_2$ with M$=$Nb, Ta)  are metallic and can exhibit superconductivity at low temperatures~\cite{Revolinsky1965,Lian2018}. Whereas TMDs with the transition metal pertaining to group VI (M$=$Mo and W~\cite{Duan2015}) have a small band-gap, that allows to overcome the zero band-gap problem present in graphene nano-sheets. Contrary to what happens to their metallic counterparts, semiconductor TMDs are stable at room temperature with no need of embedding them in a inert atmosphere~\cite{Tyagi2020}. Additionally, their band-gap is located in the near infra-red and visible radiation range, which suggests their appropriateness for solar cells fabrication~\cite{Tongay2012}. 

Among their possible applications TMD semiconductors have emerged as a good test-bed where to study valleytronic physics by taking advantage of  the broken inversion symmetry that appears between the $K^+$ and $K^-$ valleys in the monolayer limit (see, for example Refs.~\cite{Jin2018,Li2019}). Employing circularly polarized light and magnetic fields it is possible to engineer spin-selective excitations. This has been both theoretically studied~\cite{Yao2008,Xiao2012,Ulstrup2017,Xuan2020} and experimentally proved \cite{Wang2015,Nardeep2014,Lyons2019} (Also for few layered nano-structures, see for example Refs.~\cite{Gong2013,Wu2013,Arora2018,Zhao2021}).

A lot of attention has been paid to the possibility of using TMD semiconductors as gas sensors (see \cite{Liu2017,Ping2017,Zeng2018,Bernal2019,Tyagi2020}, for reviews on this topic), although up to now most of the research in this direction has been focused in  MoS$_2$ and WS$_2$ nano-sheets (see~\cite{Donarelli2018} for a comparative review of these two materials with graphene oxide). As an example, MoS$_2$ has been reported as a highly sensitive, selective, charge transfer detector for NO, NO$_2$ and NH$_3$~\cite{Li2014,Cho2015}. Notably, some efforts have been dedicated to the possibility of fabricating wearable sensors, taking advantage of the mechanical robustness and flexibility that TMDs offer~\cite{Kumar2020}. Also from  the theoretical point of view, some efforts have been committed to characterize the adsorption of light atoms~\cite{Ma2011} and molecules~\cite{Nagarajan2018,Ai2019,Zhao2014,Babar2019} on MoS$_2$, MoSe$_2$ and WS$_2$~monolayers by means of density functional theory (DFT). Similarly, molecular adsorption has been considered in Janus WSSe, MoSSe, and WSTe monolayers~\cite{Chaurasiya2019, Chaurasiya2020,Dou2021}.

Regarding the employment of  monolayer MoSe$_2$ for gas sensing, some proposals have also  appeared in the bibliography. For example the possibility of using it as a NH$_3$ detector was investigated in Refs.~\cite{Late2014,Zhang2017,Guo2019} (see also \cite{Ren2019} for  CH$_4$ detection). Interestingly, it has been shown that the size of the MoSe$_2$ nano-sheets affects the balance between the response and recovery towards gas adsorption~\cite{Zhang2017}. Besides the 2D geometry in which we focus here, other proposals appear in the bibliography such as the nano-flower CO detector that is characterized in Ref.~\cite{YANG2020127369}.

In order to improve the gas sensing performance of TMDs, different strategies have been explored. Vacancy engineering has came out as a promising research route, as it enhances the chemical activity of the surface~(see~\cite{Koos2019,Liang2021} for a review on this topic).  Alternatively, replacing a proportion of chalcogen atoms with Ge or Sb  has been demonstrated to improve the adsorption of Nitrogen containing gases~\cite{Panigrahi2019}. A similar approach, consists in doping TMDs monolayer with additional transition metal impurities. In general, the doping of the material with a metallic impurity reduces the gap giving small metallic character to the whole system, thus enhancing its conductivity, which is relevant for sensing purposes. Among other proposals~\cite{Choi2017,LIU2021127117,NI2020145911,REN2019136631,Guo2021}, in Ref.~\cite{AGUILAR2020147611}  the CO activation on the  MoS$_2$ monolayer has been studied for different metallic doping and in  Ref.~\cite{ZHANG2019930} the adsorption of H$_2$ and C$_2$H$_2$ molecules has been characterized in a Rhodium doped MoSe$_2$ sheet. Alternatively, doping with pnictogens group atoms (N, P, As) has demonstrated to enhance the adsorption energies in WSSe Janus monolayers~\cite{Abbas2018,Kaur2021}.

The usage of TMDs as catalytic surfaces is also an active research field. On one hand, it has been studied how the presence of  vacancies and metallic doping on a MoSe$_2$ nano-sheet enhances the Hydrogen evolution reaction activity~\cite{Shu2017,Lee2018,Shu2017,Jain2020}. On the other, the Oxygen evolution reaction has been studied in MoS$_2$ focusing on the pristine surface~\cite{GERMAN2020146591}, on its edges~\cite{Karmodak2021}, and also considering a reduced geometry (dots)~\cite{Mohanty2018}. Similarly, the  MoSe$_2$ and MoS$_2$ monolayers have been proposed as platforms for  N$_2$ and CO$_2$ reduction (see~\cite{Giuffredi2021} for a review). In particular, it has been revealed that the efficiency is improved at the edges of the nano-sheets~\cite{chan2014} and in the presence of chalcogen vacancies or metallic doping~\cite{Kang2019,Ye2021}.

In this work, we study the adsorption and dissociation of common diatomic molecules (H$_2$, O$_2$, CO, and NO) on the less studied MoSe$_2$ monolayer, placing special emphasis on how each of these possible adsorbates affects the electronic and magnetic properties of the MoSe$_2$ substrate. Inspired by the above mentioned works, we explicitly compare the case in which the MoSe$_2$ surface is perfectly formed (pristine case) with the one in  which a Se vacancy is present on the MoSe$_2$ monolayer (MoSe$_2$-vSe). As a previous step for the calculation of dissociation, we evaluate the most energetically favorable adsorption position across the surface for each molecule and also for its atomic components. The paper is organized as follows: In section~\ref{sec:method} we give the details about the DFT calculations that we perform. Then, in section~\ref{sec:results} we comment our results. After that, we evaluate the dissociation energy for each of the diatomic molecules on the MoSe$_2$ monolayer. The possible effect that spin-orbit coupling (SOC) may induce in the adsorption and dissociation energies that we report is investigated in Appendix~\ref{sec:spin_orbit}. Finally, we summarize our conclusions in section~\ref{sec:conclusions}.

\section{Method}
\label{sec:method}
\subsection{DFT calculations}
The DFT calculations presented here are performed with the Vienna \textit{ab initio} simulation package ({\sc vasp}) \cite{Kresse1996}. Two different approaches are used to incorporate the van der Waals (vdW) interaction in the electron exchange-correlation: the revPBE-vdW functional introduced by Dion \textit{et al.}~\cite{Dion2004} that self-consistently incorporates non-local correlation corrections and the PBE-D3 method \cite{Grime2010} that also includes the vdW correction but in a non self-consistent manner. In the latter, the exchange-correlation interaction of electrons is described by the generalized gradient approximation (GGA) in the form of Perdew-Burke-Ernzerhof (PBE) \cite{Perdew1996} and using the Becke-Johnson (BJ) damping \cite{Grime2011} to avoid strong repulsion at short distances. The ionic cores are described with the projector augmented wave (PAW) method \cite{blockl1994} implemented in {\sc vasp}~\cite{Kresse1999}. Specifically, we use the PAW potentials with 14, six, six, five, four, and one valence electrons for Mo, Se, O, N, C,and H, respectively. The Kohn-Sham states are expanded in a plane-wave basis set using as energy cutoffs: 700~eV for the systems containing C, O, and N; and 500~eV for those involving H only and for the bare monolayer. The Brillouin zone integration is performed with (cell-size adapted, see below) $\Gamma$-centered Monkhorst–Pack (MP) grids of special $\mathbf{k}$-points~\cite{Monkhorst1976}, using a Gaussian smearing of 0.1~eV for electronic state occupancies. In all the calculations, 10$^{-6}$ and 10$^{-5}$~eV are the energy converge criteria for the electronic and structural optimization, respectively. During the structural optimization all the atoms in the simulation cell are allowed to move.

The MoSe$_2$ monolayer is considered to be in the $1H$ phase. As a first step, the MoSe$_2$ hexagonal lattice constant $a$ is calculated in the monolayer primitive cell with 23~{\AA} of vacuum along the surface normal and using a 11$\times$11$\times$1 $\Gamma$-centered MP mesh of special $\mathbf{k}$-points. The values obtained with revPBE-vdW and PBE-D3 are $a$=3.386 and 3.282~{\AA}, respectively. See appendix~\ref{app:MoSe2_char} for more details. A larger 4$\times$4 surface supercell together with a 4$\times$4$\times$1 MP grid are used to study adsorption of the different gas species on pristine MoSe$_2$ and on MoSe$_2$ MoSe$_2$-vSe. This larger supercell avoids lateral interactions of the adsorbate with its images and also ensures a reliable description of an isolated Se vacancy in the MoSe$_2$-vSe surface. (See Appendix~\ref{app:MoSe2_char} for more details about MoSe$_2$ and MoSe$_2$-vSe characterization).

When needed for the discussion, the band structure and density of states (DOS) are split into atomic components. This is done in {\sc vasp} by projecting onto atomic localized orbitals \cite{Amadon2008,Karolak2011,Schler2018}. For each pair of $\mathbf{k}$-point and $n$-band index,  it projects the Kohn-Sham wave function $\phi_{n\mathbf{k}}$ onto spherical harmonics $Y_{lm}^i$ centered at each ion position $i$, such as:
\begin{equation}
    P_{n\mathbf{k},lm}^i \sim |\langle \phi_{n\mathbf{k}}|Y_{lm}^i \rangle|^2.
\end{equation}

All the structural and electron density figures were plotted with the {\sc vesta} software~\cite{vesta}, whereas the PyMatgen package~\cite{PyMatgen} was used for the band structure and DOS plots. For simplicity, only the PBE-D3 optimized structures, band structures, DOS, and charge density plots will be shown.

\subsection{Adsorption and dissociation of diatomic molecules}

Following previous works \cite{Ma2011}, we explore different possible adsorption positions across the surface. In particular, the considered adsorption sites on pristine MoSe$_2$ are: atop a Se atom ($\mathrm{T_a}$), atop a Mo atom ($\mathrm{T_b}$), on the hollow site ($\mathrm{T_h}$), midway the Mo-Se bond ($\mathrm{T_{ab}}$), and midway the $\mathrm{T_a}$ and $\mathrm{T_h}$ sites ($\mathrm{T_{ah}}$). On the MoSe$_2$-vSe surface, for simplicity, we only consider adsorption at the vacancy position ($\mathrm{T_{vSe}}$).  After all this preliminary assay, in Sec.~\ref{sec:results} only the results of the most energetically favorable configuration will be discussed.  In addition to the structural relaxation, the stability of each adsorption configuration is further confirmed by performing a vibrational mode analysis of the corresponding adsorbate, which is based on a finite-difference calculation of the Hessian matrix.

The adsorption energy $E_a$ of a molecule (or atom) $X$ on the TMD surface (MoSe$_2$ or MoSe$_2$-vSe) is calculated as,
\begin{align}
E_a (X) = E(X/\mathrm{TMD}) - E_{\mathrm{(gas)}}(X)- E(\mathrm{TMD}) \, ,
\label{eqn:adsorption_energies}
\end{align}
where $E(X/\mathrm{TMD})$ is the energy of the relaxed system composed by $X$ adsorbed on the monolayer, while $E_{\mathrm{(gas)}}(X)$ and $E(\mathrm{TMD})$ are the energies of the (optimized) isolated adsorbate and corresponding TMD monolayer, respectively.

The dissociation energy $E_d$ of a diatomic molecule $AB$ on the surface is also calculated with Eq.~\eqref{eqn:adsorption_energies}, where  $E(X/\mathrm{TMD})$ is now the energy of the two atoms $A+B$ adsorbed at separate positions on the surface. For completeness, we also calculate the dissociation energy in the limiting case in which the two atoms are separated by an infinite distance as,

\begin{align}
E_d^\infty(AB) = E_a(A)+E_a(B) + E_{d\mathrm{(gas)}}(AB) \, , 
\label{eqn:dissociation_energies2}
\end{align}
where $E_a(A)$ and $E_a(B)$ are the adsorption energy of each atom forming the molecule and $E_{d\mathrm{(gas)}}(AB)$ is the dissociation energy in gas-phase. The latter is obviously calculated as the energy difference between the isolated atoms and the isolated (optimized) molecule [$E_{d\mathrm{(gas)}}(AB) = E_{\mathrm{(gas)}}(A)+E_{\mathrm{(gas)}}(B) - E_{\mathrm{(gas)}}(AB)$]. As atomic adsorption positions for the dissociation process, we will typically consider the most stable adsorption sites that are obtained from our preliminary adsorption study for each of the relevant atomic species (H, O, C, and N). 

To characterize the molecular adsorption configuration we also compute the molecular bond length at the adsorption position d(A-B), the height $Z$ of the molecule geometrical center from the surface (defined as the average heights of the Se atoms in the topmost layer), and the polar and azimuthal angles, $\vartheta$ and $\varphi$, defining the molecular axis orientation respect to the surface normal and the lattice vector $\vec{a}$. Only d(A-B) and the height of each atom ($z_\mathrm{A}$, $z_\mathrm{B}$) are provided for the dissociated configurations.

The analysis of the charge distribution between the adsorbate $X$ and the  surface provides a deeper insight into the adsorption process. To this purpose we evaluate, depending on the system, the following charge quantities: (i) The Bader charge transfer,
\begin{equation}
 \Delta\mathcal{Q}_\mathrm{B} (X) =  \mathcal{Q}_\mathrm{B} - Z(X) \, ,
 \label{eqn:bader_charge_trans}
\end{equation}
where $\mathcal{Q}_\mathrm{B}$ is the Bader electron charge calculated with the implementation by Tang \textit{et al.}~\cite{Tang2009} and Henkelman \textit{et al.}~\cite{HENKELMAN06} and $Z(X)$ the adsorbate total atomic number (i.e., $Z(A)+Z(B)$ for the diatomic molecules); (ii) The induced electron density, 

\begin{equation}
\Delta n(\textbf{r}) = n_{X/\mathrm{TMD}}(\textbf{r}) - n_{\mathrm{TMD}}(\textbf{r}) -n_{X}(\textbf{r}) \, ,
\label{eqn:charge_diff}
\end{equation}
where $n_{\mathrm{X/TMD}}$ is the electron density distribution when $X$ is adsorbed on MoSe$_2$ (or MoSe$_2$-vSe), whereas $n_{X}$ and $n_{\mathrm{TMD}}$ are those of the isolated adsorbate and corresponding surface, respectively, which are calculated keeping the optimized 
adsorption structure in both cases; and in the case of those systems that have a finite spin magnetic moment, (iii) the spin-magnetization density,
\begin{equation}
\Delta m (\textbf{r}) = n^\uparrow(\textbf{r})-n^\downarrow(\textbf{r}),
\label{eqn:spin_diff}
\end{equation}
with $n^{i}(\textbf{r})$, $i = \uparrow, \downarrow$ the local electron density for the spin majority and minority components, respectively.

\section{Results}
\label{sec:results}

\begin{figure*}[tb!]
	\centering
	\includegraphics[width=1.0\linewidth]{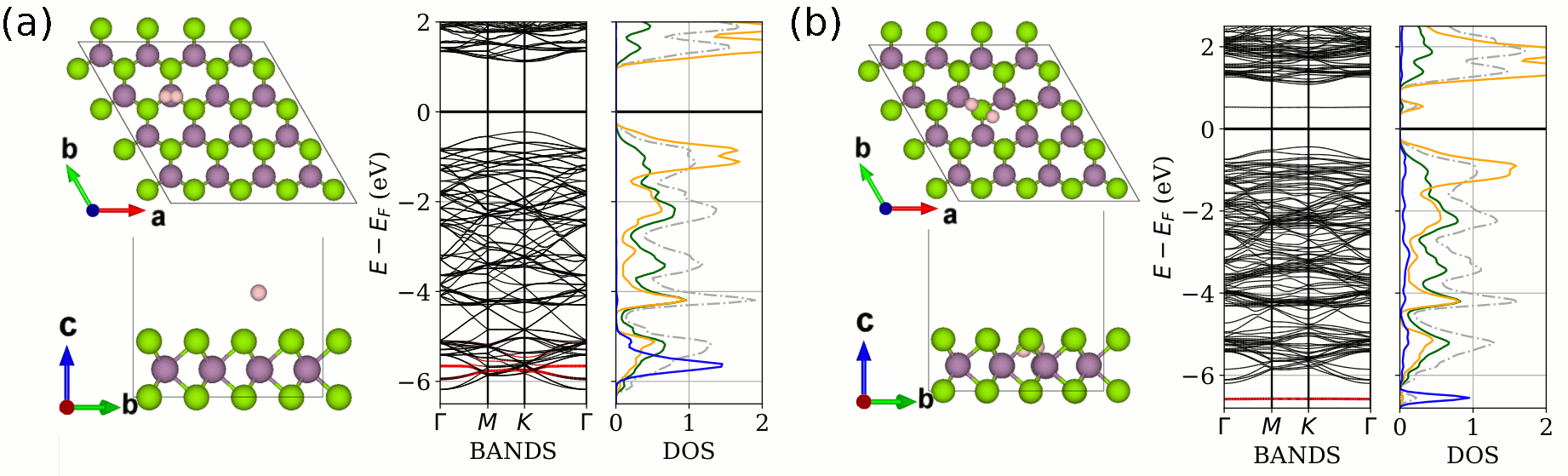}
\caption{PBE-D3 optimized structure (top and side views), band structure, and DOS for H$_2$ adsorbed at (a) T$_\mathrm{b}$ on  MoSe$_2$ and (b) the vacancy on MoSe$_2$-vSe. Se, Mo, and H atoms colored in green, purple, and pink, respectively. Bands in red correspond to projection onto the adsorbate states. Total DOS in grey, Mo-projected $d$-states in orange, Se-projected $p$-states in green, and H-projected $s$-state in blue. For clarity, all the density of states distributions are divided by the number of ions that involve.}
	\label{fig:bands_H2}
\end{figure*}

\subsection{H$_2$ adsorption and dissociation} 
\label{sec:H2}

\begin{figure}[tb]
	\centering
	\includegraphics[width=0.95\linewidth]{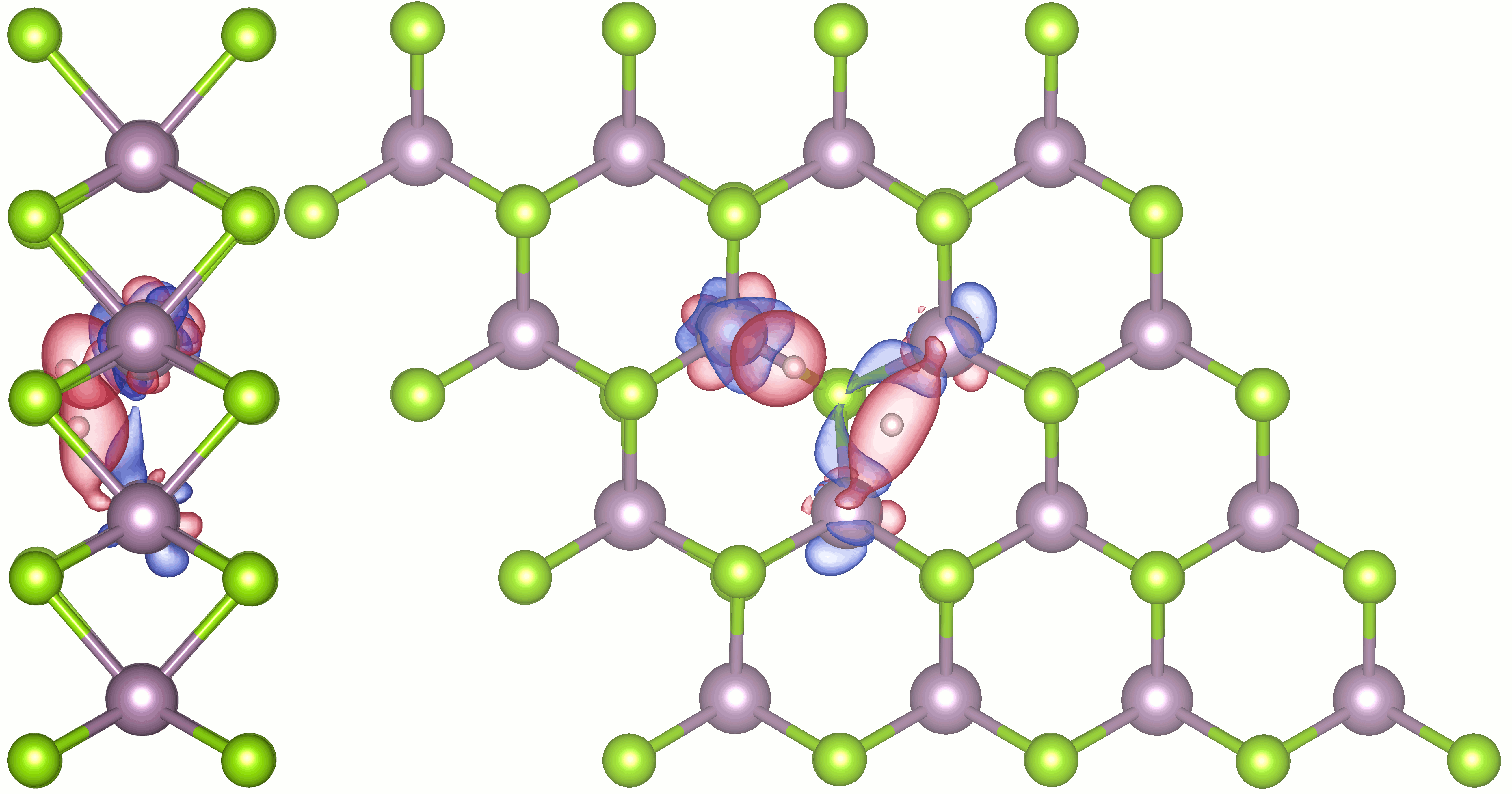}
	\caption{PBE-D3 isosurface (side and top views) of the induced electron density $\Delta n$ for H$_2$ adsorbed on MoSe$_2$-vSe. Red (Blue) surfaces indicate electron loss (gain). Atoms colored as in Fig.~\ref{fig:bands_H2}. The isovalue is 6$\times10^{-3}$ electrons/{\AA}$^{-3}$.}
		\label{fig:charge_trans_H2}
\end{figure}

\begin{table}[b]
\caption{Properties of the optimized configurations for H$_2$ adsorption on the pristine MoSe$_2$ surface ($\mathrm{T_a}$ to $\mathrm{T_{ah}}$ sites) and on MoSe$_2$-vSe ($\mathrm{T_{vSe}}$): Adsorption energy $E_a$,  bond length d(H-H), adsorption height $Z$, molecular orientation ($\vartheta$, $\varphi$), and spin-magnetic moment $|\vec{M}|$.}
\label{table:H2_MoSe2}
\begin{ruledtabular}
\begin{tabular}{ l c c c c c c c}
	\multicolumn{8}{c}{\textbf{revPBE-vdW}}\\
	\hline
	& $\mathrm{T_a}$& $\mathrm{T_b}$& $\mathrm{T_h}$& $\mathrm{T_{ab}}$& $\mathrm{T_{ah}}$& &$\mathrm{T_{vSe}}$\\  \cline{2-6}\cline{8-8}
	$E_a$(eV)	&$-$0.053 &$-$0.067& $-$0.063 &$-$0.057 &$-$0.059 &  & $-$0.181 \\
	d(H-H)(\AA) & 0.739 & 0.739 &0.740 & 0.740 &0.740 & &1.868\\
	$Z$(\AA)				& 3.75 & 3.09 &3.48 & 3.38 & 3.49 & &$-$0.54\\
	$\vartheta$  & 90$^\circ$ & 90$^\circ$ & 90$^\circ$ & 90$^\circ$ & 84$^\circ$ & & 82$^\circ$\\
	$\varphi$  & 0$^\circ$ & 0$^\circ$ & 0$^\circ$ & 60$^\circ$ & 5$^\circ$ & & 30$^\circ$\\
	$|\vec{M}|$($\mu_B$)			& 0 & 0& 0&0 & 0& &0 \\\hline
	\multicolumn{7}{c}{\vspace{0.0cm}}\\
	\multicolumn{8}{c}{\textbf{PBE-D3}}\\	
	\hline
	& $\mathrm{T_a}$& $\mathrm{T_b}$& $\mathrm{T_h}$& $\mathrm{T_{ab}}$& $\mathrm{T_{ah}}$& &$\mathrm{T_{vSe}}$\\  \cline{2-6}\cline{8-8}
	$E_a$(eV)	&$-$0.031  &$-$0.049 & $-$0.043 &$-$0.049 &$-$0.047 &  &$-$0.539 \\
	d(H-H)(\AA) & 0.750	& 0.751 & 0.750 & 0.751 & 0.751 & &1.774\\
	$Z$(\AA)				& 3.20 & 2.90& 2.96& 3.17 & 3.13 & &$-$0.58\\
	$\vartheta$  & 90$^\circ$ & 90$^\circ$ & 88$^\circ$ & 90$^\circ$ & 71$^\circ$ & & 84$^\circ$\\
	$\varphi$  & 0$^\circ$ & 0$^\circ$ &$-$7$^\circ$ & 60$^\circ$ & 22$^\circ$ & & 31$^\circ$\\
	$|\vec{M}|$($\mu_B$)			& 0 & 0 & 0 & 0& 0& & 0\\
\end{tabular}
\end{ruledtabular}
\end{table}

The experimental H$_{2\mathrm{(gas)}}$ bond length of 0.741~{\AA} \cite{NIST} is correctly reproduced by both revPBE-vdW (0.739~\AA) and PBE-D3 (0.750~\AA). Regarding the H$_{2\mathrm{(gas)}}$ dissociation energy, revPBE-vdW overestimates it by about 0.300~eV, while PBE-D3 provides a value of 4.539~eV that is in reasonable agreement with the experimental values $E_d$=4.746 and 4.787~eV. Note that, to compare it to our DFT data, the latter values are respectively obtained by adding the H$_{2\mathrm{(gas)}}$ zero point energy (ZPE) of 269 meV~\cite{Irikura2007} to the experimental dissociation energy of 4.477~eV ~\cite{Darwent1970} and   4.518~eV~\cite{Benson1965}. 

Our results for adsorption of one H$_2$ molecule at different sites on the pristine MoSe$_2$ surface and at the Se vacancy are summarized in Table~\ref{table:H2_MoSe2}. Each optimization started with the molecule oriented parallel to the surface ($\vartheta$=90$^\circ$). The structural relaxation calculations suggest that H$_2$ can adsorb in any of the five sites that were considered in the pristine MoSe$_2$ layer, being $\mathrm{T_b}$ the most energetically favorable site (although $\mathrm{T_{b}}$  and $\mathrm{T_{ab}}$ are energetically degenerated in the case of PBE-D3). The small adsorption energy of about 50--70~meV agrees with the results from the normal mode analysis at the $\mathrm{T_b}$ configuration showing that the H$_2$ molecule is weakly bounded in the perpendicular direction to the surface ($\hbar\omega\sim$~20~meV), and actually free to move parallel to the surface and to rotate around the surface normal direction. 

Regarding the structural properties, both functionals provide similar results, remaining the H$_2$ bond length practically unaltered when compared to its gas phase value. However, the revPBE-vdW predictions for the H$_2$ height from the surface $Z$ are always between 0.2 and 0.6~{\AA} larger than the PBE-D3 ones.

Figure~\ref{fig:bands_H2}(a) shows the geometrical structure, band structure, and DOS for the energetically favored $\mathrm{T_b}$ configuration. 
As observed in the H-projected band structure (red bands) and DOS (blue curve), the H$_2$-dominated $s$-like orbitals lie around 5.5~eV below the Fermi level and are rather localized in energy. The broadening of about 500 meV suggests a weak interaction with the MoSe$_2$ layer in line with the small $E_a$ value and vibrational modes. In this respect, the Bader charge analysis reveals that the charge transfer is almost negligible ($\Delta\mathcal{Q}_\mathrm{B}$=0.006~e) and, hence, it confirms that H$_2$ is physisorbed. For comparison, using the Local Density Approximation, the adsorption energy for H$_2$ adsorbed on the pristine WS$_2$ surface is 75~meV~\cite{Zhou2015}.

\begin{figure*}[tb!]
	\centering
	\includegraphics[width=1.00\linewidth]{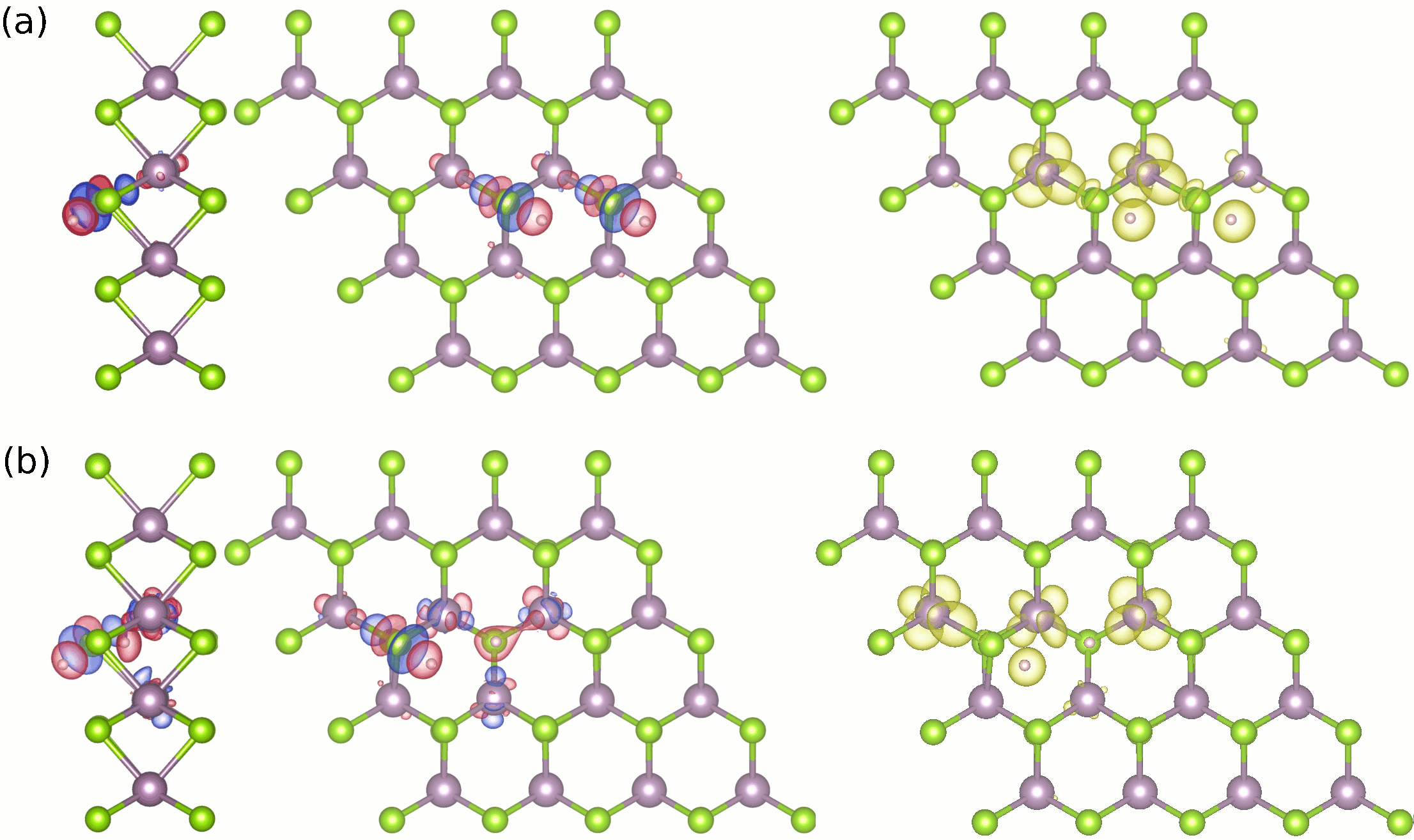}
\caption{From left to right: PBE-D3 isosurfaces of the induced electron density $\Delta n$ (side and top views) and spin-magnetization density $\Delta m$ (top view only) for H$_2$ dissociated (a) at $\mathrm{T_{ah,ah}}$ on MoSe$_2$ and (b) at $\mathrm{T_{vSe,ah}}$ on MoSe$_2$-vSe. Red (Blue) isosurfaces indicate defect (excess) of electrons and yellow (blue) isosurfaces spin majority (minority) excess. Atoms colored as in Fig.~\ref{fig:bands_H2}. An isovalue of 6$\times 10^{-3}$ electrons/{\AA}$^{-3}$ is used in all cases.}
	\label{fig:charge_dis_H}
\end{figure*}

\begin{table}[b]
 \caption{Properties of the proposed configurations for dissociative adsorption of H$_2$ on MoSe$_2$ ($\mathrm{T_{ah,ah}}$) and on MoSe$_2$-vSe ($\mathrm{T_{vSe,ah}}$) (see text for details): Dissociation energies $E_d$ and $E_d^\infty$, H-H internuclear distance d(H-H), height of each H from the surface $z_\mathrm{H}$, and system spin-magnetic moment $|\vec{M}|$.}
\label{table:dissH2_MoSe2}
\begin{ruledtabular}
\begin{tabular}{ l c c c c c}
& \multicolumn{2}{c}{\textbf{revPBE-vdW}}& & \multicolumn{2}{c}{\textbf{PBE-D3}}\\
\cline{2-3}\cline{5-6}
	&  $\mathrm{T_{ah,ah}}$& $\mathrm{T_{vSe,ah}}$ & &  $\mathrm{T_{ah,ah}}$& $\mathrm{T_{vSe,ah}}$ \\  \hline
	$E_d$(eV)	&   3.652 & 1.560 & & 3.378 & 1.256\\
	$E_d^\infty$(eV)	&   3.657 & 1.626 & & 3.368 & 1.284\\
	d(H-H)(\AA) & 3.41 & 2.84 & &  3.28 &   2.94\\
	$z_\mathrm{H}$(\AA)	& 1.33, 1.33 &$-$0.87, 1.26 & & 1.15, 1.15 &$-$0.82, 1.15 \\
	$|\vec{M}|$($\mu_B$) &2 & 2 & &2 & 2
\end{tabular}
\end{ruledtabular}
\end{table} 

On the contrary, when H$_2$ adsorbs at the Se vacancy in MoSe$_2$-vSe, strong bonds are formed. The adsorption energy in this case is significantly enhanced, being its revPBE-vdW (PBE-D3) value $E_a$=$-$0.181 ($-$0.539)~eV. Even if the discrepancy in the adsorption energy between the two functionals is considerable, the adsorption geometry is in both cases very similar (cf. last column in Table~\ref{table:H2_MoSe2}). 
At the revPBE-vdW (PBE-D3) adsorption position, the H$_2$ geometrical center is 0.54 (0.58)~{\AA} below the average MoSe$_2$-vSe surface. Remarkably, the distance between the two H atoms comes out to be as large as $\sim$1.85 (1.77)~{\AA}. This is more than twice the value of the gas-phase H$_2$ bond length [see the T$_\mathrm{vSe}$ configuration depicted in Fig.~\ref{fig:bands_H2}(b)]. In this case, the H$_2$ molecule is not parallel to the surface but slightly tilted from the surface plane, with its axis pointing to the closest Mo atom. A normal mode analysis reveals that the H$_2$ is well located at the reported position, being the energy of the lowest mode $\hbar\omega\sim$~65~meV. The stretch in the H-H bond length that we find here is in agreement with the one obtained with the PBE-D2 method in Ref.~\cite{Shu2017}, where an adsorption energy of $-$0.621 eV is also reported.

The corresponding band structure and DOS are also shown in Fig.~\ref{fig:bands_H2}(b). Upon adsorption of H$_2$ in the Se vacancy, the MoSe$_2$-vSe in-gap states remain unoccupied, although the direct band-gap (between the valence band maximum and the lowest in-gap) decreases by 80~meV. The H-projected DOS exhibits a narrow peak at about $-6.5$~eV that, being readily associated to the flat red bands, is not expected to contribute much to the surface-molecule bonding. We additionally observe a large spread of the projected DOS along the valence MoSe$_2$ states that overlaps with the DOS projected onto $p$- and $d$-states of Selenium and Molybdenum, respectively. The latter may hint to a sizeable charge transfer between the adsorbate and the MoSe$_2$ layer that is confirmed by the Bader charge analysis that gives $\Delta\mathcal{Q}_\mathrm{B} =$ 0.54~$e$. The $\Delta n$ isosurface of Fig.~\ref{fig:charge_trans_H2} shows that the charge transfer comes mainly from the three closest Mo atoms.

Regarding the atomic adsorption of H, which is needed to evaluate the H$_2$ dissociation energy, the $\mathrm{T_{ah}}$ site in the pristine MoSe$_2$  surface is the most energetically favorable adsorption position, 
being the revPBE-vdW (PBE-D3) prediction for the adsorption energy $E_{a}$(H)= $-$0.689 ($-$0.580)~eV. However, in the case of MoSe$_2$-vSe, H adsorbs preferentially in the Se vacancy with a revPBE-vdW (PBE-D3) energy of  $-$2.715 ($-$2.674)~eV (see Appendix~\ref{app:at_ener} for all the details). In view of these results, our study of the dissociation process in MoSe$_2$ is restricted to the case in which the two H atoms end up adsorbed at two adjacent $\mathrm{T_{ah}}$ positions; whereas in MoSe$_2$-vSe, one H is adsorbed in the vacancy and the other H in the adjacent $\mathrm{T_{ah}}$ site ($\mathrm{T_{vSe,ah}}$).

Our results for the dissociative adsorption of H$_2$ on MoSe$_2$ are summarized in Table~\ref{table:dissH2_MoSe2}. Even if the gas-phase dissociation energy is reduced by about 1.3~eV on the surface, the process remains highly endothermic ($E_d$=3.652 and 3.378~eV with revPBE-vdW and PBE-D3, respectively). For each functional, the $E_d$ and $E_d^\infty$ values are very similar, reflecting the lack of interaction between H atoms separated by one lattice constant. It is worth noting that the dissociative adsorption of H$_2$ induces a spin magnetization of 2~$\mu_B$ in the system. Figure~\ref{fig:charge_dis_H}(a) shows isosurfaces of the induced electron density  $\Delta n(\textbf{r})$ and spin-magnetization density $\Delta  m(\textbf{r})$ for H$_2$ dissociatively adsorbed on MoSe$_2$. Each H atom binds to its closest Se, causing a charge rearrangement that is very localized along the corresponding H-Se-Mo line. Remarkably, although the total spin-magnetic moment corresponds to that of the two isolated H atoms, the spin-magnetization density $\Delta  m(\textbf{r})$ is located both around the H atoms and the Mo atoms involved in the mentioned charge rearrangement. Also a spin-magnetization of 1~$\mu_\mathrm{B}$ has been obtained for atomic adsorption of H on MoSe$_2$, MoTe$_2$, and WS$_2$ \cite{Ma2011}, and on MoS$_2$ \cite{he10}. The possibility of magnetizing locally the material by adsorbing H atoms has already been predicted and observed in the case of Graphene \cite{yazyev07, boukhavalov08, Gonzalez2016}. 

As seen in Table~\ref{table:dissH2_MoSe2}, the dissociation energy is further reduced when a Se vacancy is present in the MoSe$_2$ sheet, but the process is also endothermic in this case. The revPBE-vdW (PBE-D3) dissociation energy is $E_d=$~1.560 (1.256)~eV. These values increase slightly (i.e., they are more endothermic), if one considers that the two H atoms diffuse to positions that are far from each other (see the corresponding $E_d^\infty$ values). Figure~\ref{fig:charge_dis_H}(b) shows isosurfaces of the induced electron density  $\Delta n(\textbf{r})$ and spin-magnetization density $\Delta  m(\textbf{r})$ for this configuration. The perturbation caused by the  H adsorbed at the $\mathrm{T_{ah}}$ position is very similar to the one created in the pristine surface [cf. Figs.~\ref{fig:charge_dis_H}(a) and (b)]. 
The H atom adsorbed at the vacancy bonds to the three surrounding Mo atoms, but it only induces spin magnetization in two of them due to the presence of the other H. (We have confirmed that the spin magnetization is equally distributed among the three Mo if we remove the H at $\mathrm{T_{ah}}$  from the MoSe$_2$-vSe monolayer.) 

Interestingly, our calculations show that H$_2$ dissociation on MoSe$_2$-vSe becomes exothermic when the two H atoms are adsorbed in two different and isolated Se vacancies, being the revPBE-vdW (PBE-D3) dissociation energy $E_{d}(\mathrm{H}_2)= -0.400$ ($-0.809$)~eV. Even if a proper study of this particular case would require to consider the energetics of the diffusion process, which is beyond the scope of this study, this result suggests that dissociation might be possible with a larger vacancy concentration. Therefore, for completeness, we have evaluated the dissociation energy for H$_2$ in the case in which two Se vacancies are present at neighbor positions. The process becomes exothermic, being the revPBE-vdW (PBE-D3) prediction $E_{d}(\mathrm{H}_2)$= $-0.560$ ($-1.008$)~eV.

DFT-D2 calculations on H$_2$ adsorption at the S vacancy of the MoS$_2$ monolayer show that the dissociative adsorption on that surface is $\sim$~300~meV more favourable than molecular adsorption \cite{Li2016}. This finding is in sharp contrast to what we find here for H$_2$ adsorption on MoSe$_2$-vSe, where the dissociation process is 1.7~eV more endothermic than the molecular adsorption.

\subsection{O$_2$ adsorption and dissociation} 
\label{subsect:O2}

\begin{figure*}[tb!]
\includegraphics[width=1.0\linewidth]{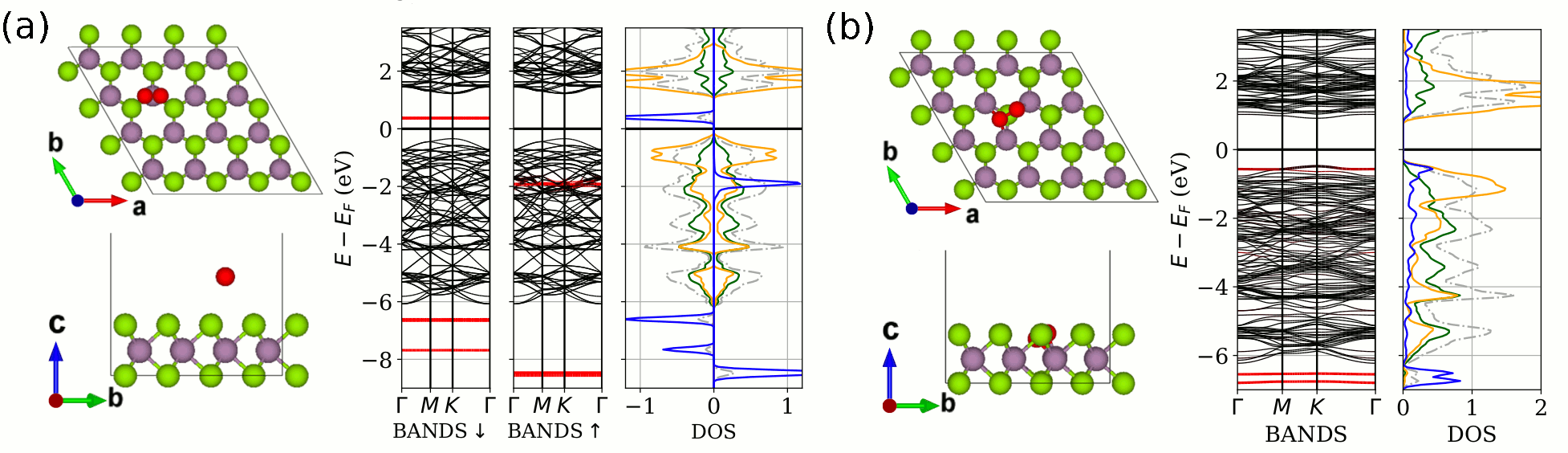}
\caption{PBE-D3 optimized structure (top and side views), band structure, and DOS for O$_2$ adsorbed at (a) T$_\mathrm{b}$ on  MoSe$_2$ and (b) the vacancy on MoSe$_2$-vSe. Se, Mo, and O atoms colored in green, purple, and red, respectively. Bands in red correspond to projection onto the adsorbate states. Total DOS in grey, Mo-projected $d$-states in orange, Se-projected $p$-states in green, and O-projected $p$-states in blue. In (a), spin-up (spin-down) bands and positive (negative) DOS refer to the spin majority (minority) components. For clarity, all the density of states distributions are divided by the number of ions that involve, except for O that is divided by two.}
\label{fig:bands_O2}
\end{figure*}

\begin{table}
 \caption{Properties of the optimized configurations for O$_2$ adsorption on the pristine MoSe$_2$ surface ($\mathrm{T_a}$ to $\mathrm{T_{ah}}$ sites) and on MoSe$_2$-vSe ($\mathrm{T_{vSe}}$): Adsorption energy $E_a$, bond length d(O-O), adsorption height $Z$, molecular orientation ($\vartheta$, $\varphi$), and spin-magnetic moment $|\vec{M}|$.}
\label{table:O2_MoSe2}
\begin{ruledtabular}
\begin{tabular}{ l c c c c c c c}
	\multicolumn{8}{c}{\textbf{revPBE-vdW}}\\
	\hline

	& $\mathrm{T_a}$& $\mathrm{T_b}$& $\mathrm{T_h}$& $\mathrm{T_{ab}}$& $\mathrm{T_{ah}}$& &$\mathrm{T_{vSe}}$\\  \cline{2-6}\cline{8-8}
	$E_a$(eV)	& $-$0.079& $-$0.110& $-$0.104& $-$0.097 &$-$0.102 &  & $-$2.394 \\
	d(O-O)(\AA) & 1.232 & 1.234	&1.233	& 1.231&1.232 & & 1.478\\
	$Z$(\AA)				& 3.40 	& 3.21	&3.20	& 3.14&3.22 & & $-$0.12\\
	$\vartheta$  & 90$^\circ$ & 90$^\circ$ & 84$^\circ$ & 89$^\circ$ & 76$^\circ$ & & 74$^\circ$\\
	$\varphi$  & 0$^\circ$ & 0$^\circ$ & 15$^\circ$ & 60$^\circ$ & 14$^\circ$ & &$-$33$^\circ$\\
	$|\vec{M}|$($\mu_B$)			& 2 &2 	&2 		& 2 & 2 & & 0 \\\hline
	\multicolumn{7}{c}{\vspace{0.0cm}}\\
	\multicolumn{8}{c}{\textbf{PBE-D3}}\\	
	\hline
	& $\mathrm{T_a}$& $\mathrm{T_b}$& $\mathrm{T_h}$& $\mathrm{T_{ab}}$& $\mathrm{T_{ah}}$& &$\mathrm{T_{vSe}}$\\  \cline{2-6}\cline{8-8}
	$E_a$(eV)	& $-$0.085 & $-$0.172& $-$0.101 &0.107 & $-$0.094& &$-$2.266\\
	d(O-O)(\AA) & 1.212 & 1.220& 1.219& 1.220&1.219 & & 1.446\\
	$Z$(\AA)				& 3.48 & 3.22 & 3.22 & 3.20&3.39 & & $-$0.15\\
	$\vartheta$  & 90$^\circ$ & 90$^\circ$ & 89$^\circ$ & 90$^\circ$ & 83$^\circ$ & & 73$^\circ$\\
	$\varphi$  & 3$^\circ$ &$-$1$^\circ$ & 1$^\circ$ & 60$^\circ$ &$-$7$^\circ$ & &$-$34$^\circ$\\
	$|\vec{M}|$($\mu_B$)			& 2 & 2& 2& 2 &2 & & 0 
\end{tabular}
\end{ruledtabular}
\end{table} 

Except for a slight stretching, the experimental O$_{2(\mathrm{gas})}$ bond length in gas-phase (1.208~{\AA})~\cite{NIST} is well reproduced by revPBE-vdW (1.232~{\AA}) and PBE-D3 (1.219~{\AA}). In contrast, the calculated revPBE-vdW (6.349~eV) and PBE-D3 (6.886~eV) O$_{2(\mathrm{gas})}$ dissociation energies overestimate the experimental values, $E_d=$~5.213 and 5.267~eV, that are obtained after adding the O$_{2(\mathrm{gas})}$ ZPE of 98~meV~\cite{Irikura2007} to the gas-phase dissociation energy of 5.115~eV~\cite{Darwent1970} and 5.169~eV~\cite{Benson1965}.

Regarding the adsorption of O$_2$ on the MoSe$_2$ monolayer, Table~\ref{table:O2_MoSe2} summarizes our results for O$_2$ adsorption at different sites. In all cases, the O$_2$ molecular axis was initially oriented parallel to the surface.  Similarly to what we find for H$_2$, the five studied configurations remain as possible O$_2$ adsorption sites after the structural relaxation. Energetically, the most favorable adsorption configuration is $\mathrm{T_b}$, with a revPBE-vdW (PBE-D3) adsorption energy $E_{a}(\mathrm{O}_2)=-$~0.110 ($-$0.172)~eV. The molecule lies parallel to the surface at a distance of about 3~{\AA} and preserves its gas-phase bond length. Our normal mode analysis for this site reveals that O$_2$ is weakly bounded in the perpendicular direction to the surface ($\hbar\omega\sim$~20~meV), 
and that the molecule will be able to perform both small displacements parallel to the surface and rotations around the surface normal under small perturbations ($\hbar\omega\le$~13~meV).

The optimized geometry for O$_2$ adsorbed at the $\mathrm{T_b}$ site on MoSe$_2$, as well as the spin-resolved band structure and spin-resolved DOS are shown in Fig.~\ref{fig:bands_O2}(a).  
The lack of dispersion in the O$_2$-dominated orbitals (flat red bands and the corresponding sharp blue peaks in the O-projected DOS) suggests a weak interaction between the molecule and the surface. The minor Bader charge transfer of 0.02~$e$ associated with the adsorbed O$_2$ confirms the physisorption character of the interaction. The induced electron density $\Delta n(\textbf{r})$ and spin-magnetization density $\Delta  m(\textbf{r})$ (not shown) also corroborate that there is only a very minor charge rearrangement on O$_2$ that conserves the gas-phase spin magnetization.
Let us remark that the O$_2$ adsorption energy in the pristine MoSe$_2$ surface is similar to the ones obtained for adsorption in other TMD monolayers by other authors. See for example Bui \textit{et al}.~\cite{Bui_2015} and Zhou~\textit{et al}.~\cite{Zhou2015} for a study on WS$_2$ and Zhao \textit{et al}. \cite{Zhao2014} for a study in MoS$_2$.

\begin{figure}[tb]
	\centering
	\includegraphics[width=0.95\linewidth]{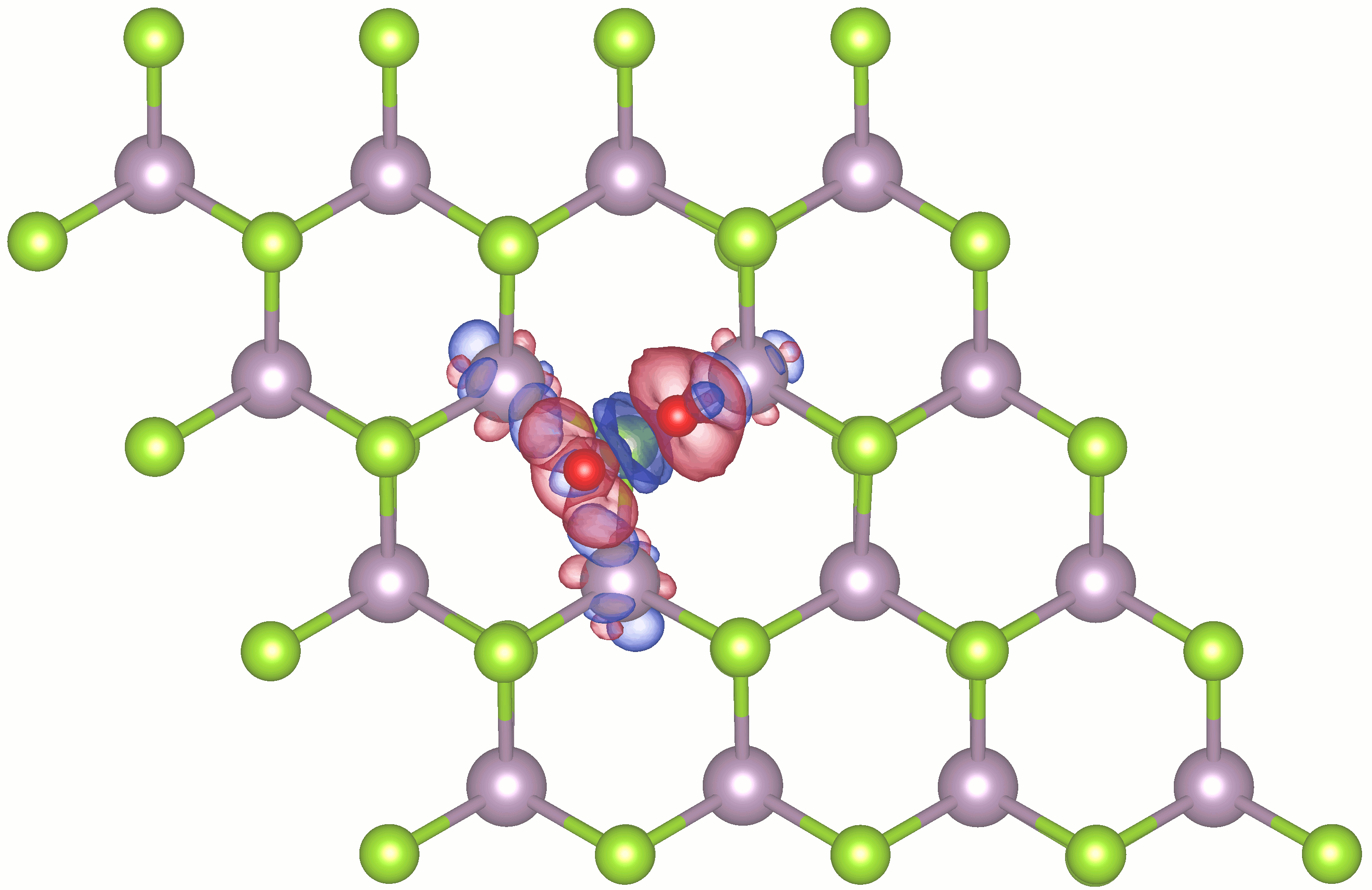}
	\caption{PBE-D3 isosurface of the induced electron density $\Delta n$ for O$_2$ adsorbed on MoSe$_2$-vSe. Red (Blue) surfaces indicate electron loss (gain). Atoms colored as in Fig.~\ref{fig:bands_O2}. The isovalue is 6$\times10^{-3}$ electrons/{\AA}$^{-3}$.}
	\label{fig:charge_trans_O2}
\end{figure}
\begin{figure}[tb]
	\centering
	\includegraphics[width=0.95\linewidth]{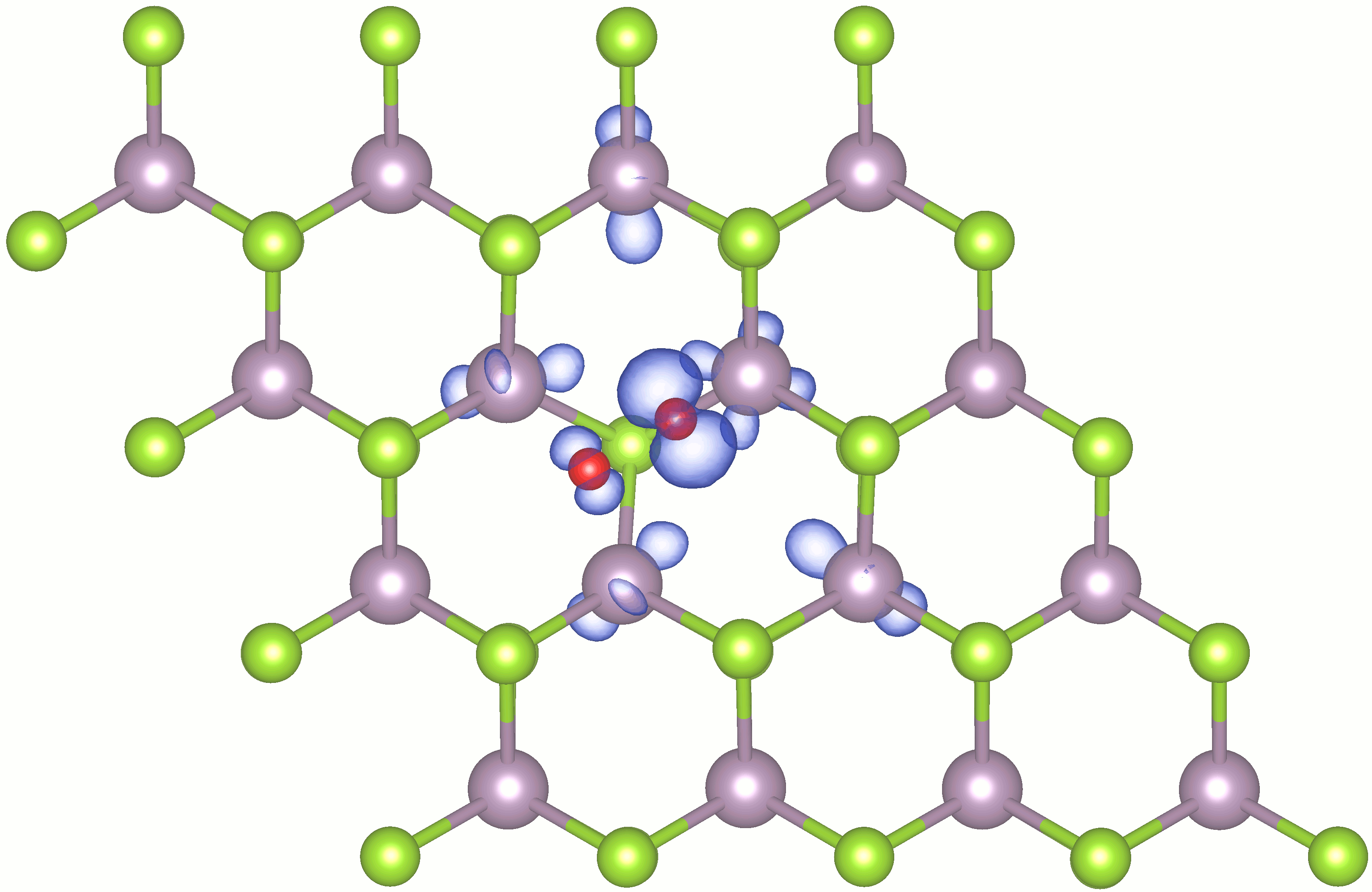}
	\caption{PBE-D3 partial electron density corresponding to the two O- and Mo-contributed quasi-degenerated bands that appear below the Fermi level upon O$_2$ adsorption at the Se vacancy.  Atoms colored as in Fig.~\ref{fig:bands_O2}. The isosurface value is 6$\times 10^{-3}$ electrons/{\AA}$^{-3}$. }
	\label{fig:parcharge_O2}
\end{figure}

Table~\ref{table:O2_MoSe2} shows that the O$_2$ adsorption energy in a Se vacancy ($\mathrm{T_{vSe}}$) is one order of magnitude larger than in the pristine surface, being the revPBE-vdW  and PBE-D3 values $-$2.394 and $-$2.266~eV, respectively. The normal mode analysis performed for this configuration shows that adsorption at the vacancy is stable at low temperatures, being the lowest energy mode $\hbar\omega\sim$~25~meV. In particular, for modes involving desorption,  $\hbar\omega\ge$~40~meV. Noticeably, the total spin-magnetic moment in this case vanishes as a consequence of a significant charge transfer from the surface to the molecule that, according to the Bader charge analysis is 1.11~$e$. Inspection of the spin-magnetization density confirms the absence of possible regions with opposite spin that could contribute to the zero spin magnetization of the system. 
The induced electron density $\Delta n(\textbf{r})$ in  Fig.~\ref{fig:charge_trans_O2} shows that O$_2$ at the $\mathrm{T_{vSe}}$ site decreases its electron density around its two atoms and gains it in the bond. 

The O$_2$/MoSe$_2$-vSe band structure in Fig.~\ref{fig:bands_O2}(b) resembles 
that of H$_2$ adsorbed at the Se vacancy [Fig.~\ref{fig:bands_H2}(b)] as the DOS projected onto the adsorbate orbitals spreads over the MoSe$_2$ valence band. However, the large charge transfer experienced by the adsorbed O$_2$ causes an important modification of the band structure that is not observed in H$_2$/MoSe$_2$-vSe. The MoSe$_2$-vSe in-gap states disappear and a O$_2$ dominated flat band appears near the valence band maximum. As a result, the system direct band-gap is 1.40~eV, which is quite similar to the corresponding value of the pristine MoSe$_2$ monolayer. The partial charge associated with the above mentioned states is plotted in Fig.~\ref{fig:parcharge_O2}. 

Adsorption of O$_2$ at a chalcogen vacancy in different TMD monolayers (MX$_2$ with M$=$W, Mo and X$=$S, Se, Te) has been studied by means of the DFT-D2 method~\cite{Chen2015,Li2016,Liu2015}. Overall, the reported adsorption energies are between 2--2.5~eV and the Bader charge captured by the adsorbed O$_2$ is around 1~$e$ in all cases. In particular, for MoSe$_2$~\cite{Liu2015}, the reported O$_2$ adsorption energy, O-O bond length, and Bader charge transfer, are 2.2~eV, 1.45~{\AA}, and 1.0~$e$, respectively,  in good agreement with the results presented here. The band structures presented in these works exhibit similar phenomenology to that of Fig.~\ref{fig:bands_O2}(b): After O$_2$ adsorption at the chalcogen vacancy the ingap states related to the presence of a chalcogen defect are removed and new flat bands are formed at a lower energy. Similar results have been found in studies with other functionals~\cite{Zhao2017,Wang2021}. 

Regarding the Oxygen atomic adsorption, our analysis in Appendix~\ref{app:at_ener} shows that  $\mathrm{T_a}$ on MoSe$_2$ and $\mathrm{T_{vSe}}$ on MoSe$_2$-vSe are the energetically favored adsorption sites. Thus, as dissociate adsorption configurations, we consider that the O atoms adsorb on the pristine surface at two $\mathrm{T_a}$ positions separated by a lattice constant ($\mathrm{T_{a,a}}$), whereas on MoSe$_2$-vSe, one O is in the vacancy and the other O at the nearest $\mathrm{T_a}$ ($\mathrm{T_{vSe,a}}$). The results are shown in Table~\ref{table:dissO2_MoSe2}.
\begin{table}
\caption{Properties of the proposed configurations for dissociative adsorption of O$_2$ on MoSe$_2$ ($\mathrm{T_{a,a}}$) and on MoSe$_2$-vSe ($\mathrm{T_{vSe,a}}$)  (see text for details): Dissociation energies $E_d$ and $E_d^\infty$, O-O internuclear distance d(O-O), height of each O from the surface $z_\mathrm{O}$, and system spin-magnetic moment $|\vec{M}|$.}
\label{table:dissO2_MoSe2}
\begin{ruledtabular}
\begin{tabular}{ l c c c c c}
	& \multicolumn{2}{c}{\textbf{revPBE-vdW}}& & \multicolumn{2}{c}{\textbf{PBE-D3}}\\
	\hline
    & T$_{a,a}$& $\mathrm{T_{vSe,a}}$ & & T$_{a,a}$& $\mathrm{T_{vSe,a}}$\\  \hline
	$E_d$ (eV) & 1.222  &$-$3.268&	& 1.039 &$-$3.947\\
	$E_d^\infty$ (eV) & 0.605 & $-$3.820&	& 0.426 &$-$3.958\\
	d(O-O)(\AA)	& 3.62 & 4.01 & &  3.54 & 3.87  \\
	$z_\mathrm{O}$(\AA) & 1.59, 1.59 & $-$0.62, 1.57 & & 1.57, 1.57 & $-$0.65,  1.52\\
	$|\vec{M}|$($\mu_B$)	 	 & 0  & 0 & &  0 & 0
\end{tabular}
\end{ruledtabular}
\end{table}
\begin{figure*}
	\includegraphics[width=1.00\linewidth]{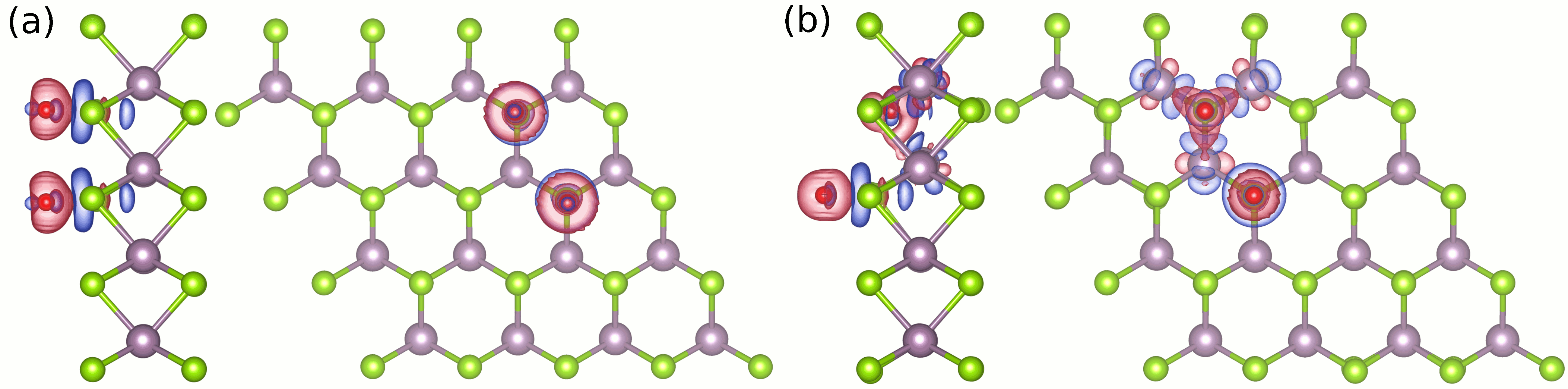}
\caption{PBE-D3 isosurfaces of the induced electron density $\Delta n$ (side and top views) for O$_2$ dissociated at (a) $\mathrm{T_{a,a}}$ on MoSe$_2$ and (b) $\mathrm{T_{vSe,a}}$ on MoSe$_2$-vSe. Red (Blue) isosurfaces indicate defect (excess) of electrons. Atoms colored as in Fig.~\ref{fig:bands_O2}. An isovalue of 6$\times 10^{-3}$ electrons/{\AA}$^{-3}$ is used in all cases.}
\label{fig:charge_dis_O}
\end{figure*}

Dissociation on the pristine surface is still endothermic, although the revPBE-vdW and PBE-D3 dissociation energies ($E_d$=1.222 and 1.039~eV, respectively) are significantly smaller than the gas-phase value. The fact that $E_d$ is about 0.6~eV smaller than $E_d^\infty$ reveals the existence of a non-negligible vdW interaction between the adsorbed O atoms in spite of being separated by 3.5--3.6~{\AA}. In this situation the Bader charge transfer associated with each O is 0.86~$e$. The excess of charge between each O and its closest Se that is observed in the induced electron density of Fig.~\ref{fig:charge_dis_O}(a) remarks the strong bonds that they form. Since the charge rearrangement is very localized along each of these bonds, we interpret the remanent interaction between the adsorbed O as a dipole-dipole interaction between the two O-Se compounds.

At variance with our results for H$_2$ dissociation, a single Se vacancy suffices to facilitate the dissociation of O$_2$. Table~\ref{table:dissO2_MoSe2} shows that the revPBE-vdW and PBE-D3 predictions for the dissociation energy at $\mathrm{T_{vSe,a}}$ are $E_{d}\sim -3.3$ and $-$3.9~eV, respectively. The revPBE-vdW $E_d^\infty$ value suggests a non-negligible interaction between the dissociated atoms, but not the PBE-D3 one. The induced electron density of Fig.~\ref{fig:charge_dis_O}(b) shows the strong bonds that the O at the vacancy forms with the three closest Mo atoms. The other O binds to the Se below, being the corresponding $\Delta n$ quite similar to what we observe for adsorption on the pristine surface [cf.  Fig.~\ref{fig:charge_dis_O}(a)].

For completeness, we have evaluated the dissociation energy of O$_2$ considering that two Se vacancies are present at neighbor lattice positions. In this situation the process is even more exothermic with a revPBE-vdW (PBE-D3) prediction $E_d = -8.090$ ($-8.219$)~eV. Similar values are obtained for the case in which the two O atoms are adsorbed at two different isolated Se vacancies $E_d^\infty\sim-8.3$~eV.

The effect that different S vacancy concentrations have in the dissociation of O$_2$ on the MoS$_2$ surface has been characterize in Ref.~\cite{Zhou2021}. Specifically, the process  becomes more favorable as the concentration of vacancies increases from 3.13\% to 12.5\%. These results are in accordance to what we find in our calculations where one and two vacancies are present in the 4$\times$4 supercell, corresponding to a concentration of vacancies  of 3.13\% and 6.25\%, respectively. 

\subsection{CO adsorption and dissociation} 
\label{{subsect:CO}}

The CO$_\mathrm{(gas)}$ bond length calculated with revPBE-vdW and PBE-D3 is 1.137 and 1.135~{\AA}, respectively. Both values reproduce the experimental value of 1.128~{\AA} \cite{NIST} within a relative error of 1\%. The revPBE-vdW (PBE-D3) dissociation energy of CO$_\mathrm{(gas)}$, $E_d$=11.68 (12.10)~eV, overestimates the corresponding experimental values, $E_d$=11.243~eV and $E_d$=11.274~eV, that, for comparative purposes, we obtain by adding the CO$_\mathrm{(gas)}$ ZPE of 134~meV~\cite{Irikura2007} to the experimental dissociation energies, 11.109~eV~\cite{Darwent1970} and 11.140~eV~\cite{Benson1965}, respectively. 

\begin{figure*}
	\centering
	\includegraphics[width=1.0\linewidth]{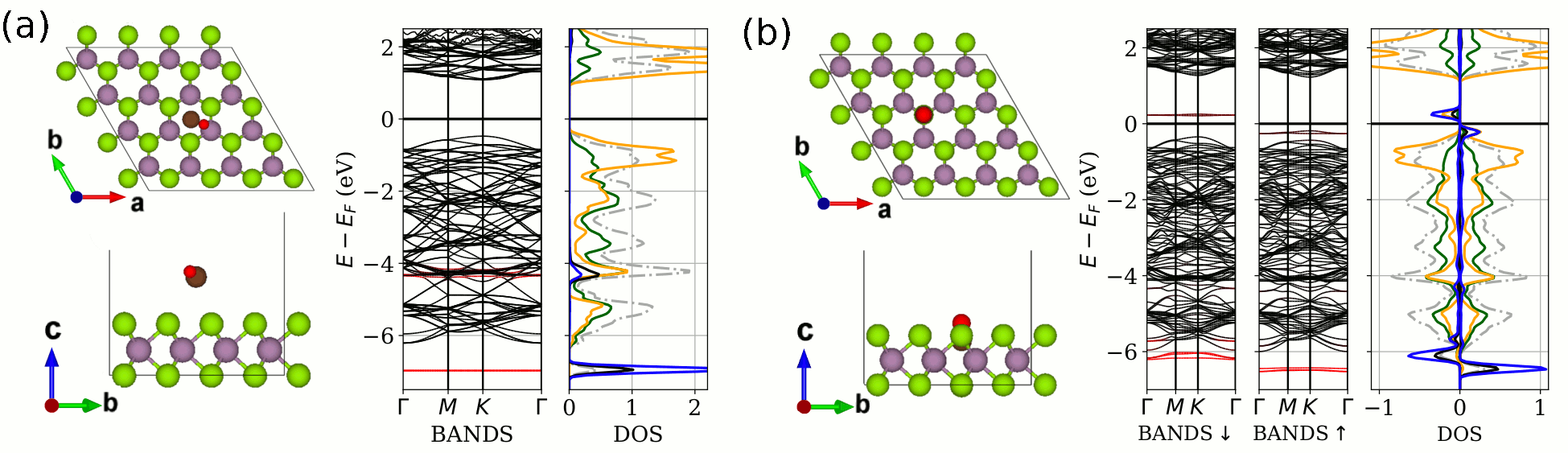}
\caption{PBE-D3 optimized structure (top and side views), band structure, and DOS for CO adsorbed at (a) T$_\mathrm{h}$ on  MoSe$_2$ and (b) the vacancy on MoSe$_2$-vSe. Se, Mo, C, and O atoms colored in green, purple, brown, and red, respectively. Bands in red correspond to projection onto the adsorbate states. Total DOS in grey, Mo-projected $d$-states in orange, Se-projected, C-projected, and O-projected $p$-states in green, black, and blue, respectively. In (b), spin-up (spin-down) bands and positive (negative) DOS refer to the spin majority (minority) components. For clarity, all the density of states distributions are divided by the number of ions that involve, except for C and O that are divided by two and three, respectively.}
	\label{fig:bands_CO}
\end{figure*}

\begin{table}
\caption{Properties of the optimized configurations for CO adsorption on the pristine MoSe$_2$ surface ($\mathrm{T_a}$ to $\mathrm{T_{ah}}$ sites) and on MoSe$_2$-vSe ($\mathrm{T_{vSe}}$): Adsorption energy $E_a$, bond length d(C-O), adsorption height $Z$, molecular orientation ($\vartheta$, $\varphi$), and spin-magnetic moment $|\vec{M}|$.}
\label{table:CO_MoSe2}
\begin{ruledtabular}
\begin{tabular}{ l c c c c c c c}
	\multicolumn{8}{c}{\textbf{revPBE-vdW}}\\
	\hline

	& $\mathrm{T_a}$& $\mathrm{T_b}$& $\mathrm{T_h}$& $\mathrm{T_{ab}}$& $\mathrm{T_{ah}}$& &$\mathrm{T_{vSe}}$\\  \cline{2-6}\cline{8-8}
	$E_a$(eV)	& $-$0.133 & $-$0.122	& $-$0.143 	& $-$0.127 & $-$0.119  & & $-$0.871\\
	d(C-O)(\AA)				&  1.137 &  1.137 	&  1.137 	& 1.137 & 1.138 & &1.208\\
	$Z$({\AA})			&  3.88 & 	3.93	& 3.51	& 4.13 & 4.24 & & 0.25\\
	$\vartheta$  & 65$^\circ$ & 14$^\circ$ & 82$^\circ$ & 36$^\circ$ & 28$^\circ$ & & 0$^\circ$\\
	$\varphi$  & 38$^\circ$ &$-$81$^\circ$ &$-$27$^\circ$ &$-$61$^\circ$ &$-$67$^\circ$ & & 0$^\circ$\\
	$|\vec{M}|$($\mu_B$)	 			&  0 	& 	0 		& 0 		&0 &0 & & 2 \\
	\hline
	\multicolumn{7}{c}{\vspace{0.0cm}}\\
	\multicolumn{8}{c}{\textbf{PBE-D3}}\\	
	\hline
	& $\mathrm{T_a}$& $\mathrm{T_b}$& $\mathrm{T_h}$& $\mathrm{T_{ab}}$& $\mathrm{T_{ah}}$& &$\mathrm{T_{vSe}}$\\  \cline{2-6}\cline{8-8}
	$E_a$(eV)	&$-$0.124 & $-$0.123 & $-$0.145 	&$-$0.120 &$-$0.103  & & $-$1.317\\
	d(C-O)({\AA})		& 1.137 & 1.134	& 1.136	&1.135	& 1.136	& &1.210 \\
	$Z$({\AA})			& 3.71  & 3.70	& 3.34	& 3.94	& 4.09	& &0.21 \\
	$\vartheta$  & 58$^\circ$ & 6$^\circ$ & 45$^\circ$ & 28$^\circ$ & 17$^\circ$ & & 0$^\circ$\\
	$\varphi$  &$-$43$^\circ$ &$-$85$^\circ$ &$-$30$^\circ$ &$-$66$^\circ$ &$-$74 $^\circ$ & & 0$^\circ$\\
	$|\vec{M}|$($\mu_B$)	 	 & 0 & 0 & 0	& 0&0 & &2 
\end{tabular}
\end{ruledtabular}
\end{table} 
Table~\ref{table:CO_MoSe2} shows our results for adsorption of one CO molecule at the five considered sites in the MoSe$_2$ surface and at the Se vacancy in MoSe$_2$-vSe. All the relaxation calculations start with the expected CO orientation for adsorption at surfaces, i.e., perpendicular to the surface being C the closest atom to the surface. Starting with the results for MoSe$_2$, both functionals provide compatible predictions for all the properties presented in Table~\ref{table:CO_MoSe2} and agree that the most energetically favorable position is $\mathrm{T_h}$, with $E_a\sim -$0.145~eV. In this configuration, CO adsorbs at $Z\sim$~3.2~{\AA}, keeping the gas-phase bond length, and it is slightly tilted respect to the surface normal, with the C atom atop $\mathrm{T_h}$ and the O atom displaced towards the nearest $\mathrm{T_b}$. The optimized structure is shown in Fig.~\ref{fig:bands_CO}(a) together with the corresponding band structure and DOS. The fact that CO is physisorbed is reflected in the rather flat bands associated with the outer CO levels that lie about 4.4~eV below the Fermi level and in the small charge transfer that is obtained with a Bader charge analysis ($\Delta\mathcal{Q}_\mathrm{B} =$~0.015~$e$). 
By analysing the induced electron density (not shown), we observe that there is only a small charge redistribution inside the CO and the closest surface atoms. A normal mode analysis shows that the molecule is only weakly bounded to the surface ($\hbar\omega\sim$~15~meV) and that is almost free to  displace in the plane parallel to the surface and to rotate around the surface normal direction ($\hbar\omega<$ 15 meV). However, perturbations involving a change on the molecule tilting angle will require a larger energy supply~($\hbar\omega>$ 20 meV).

\begin{figure}
	\centering
	\includegraphics[width=1.00\linewidth]{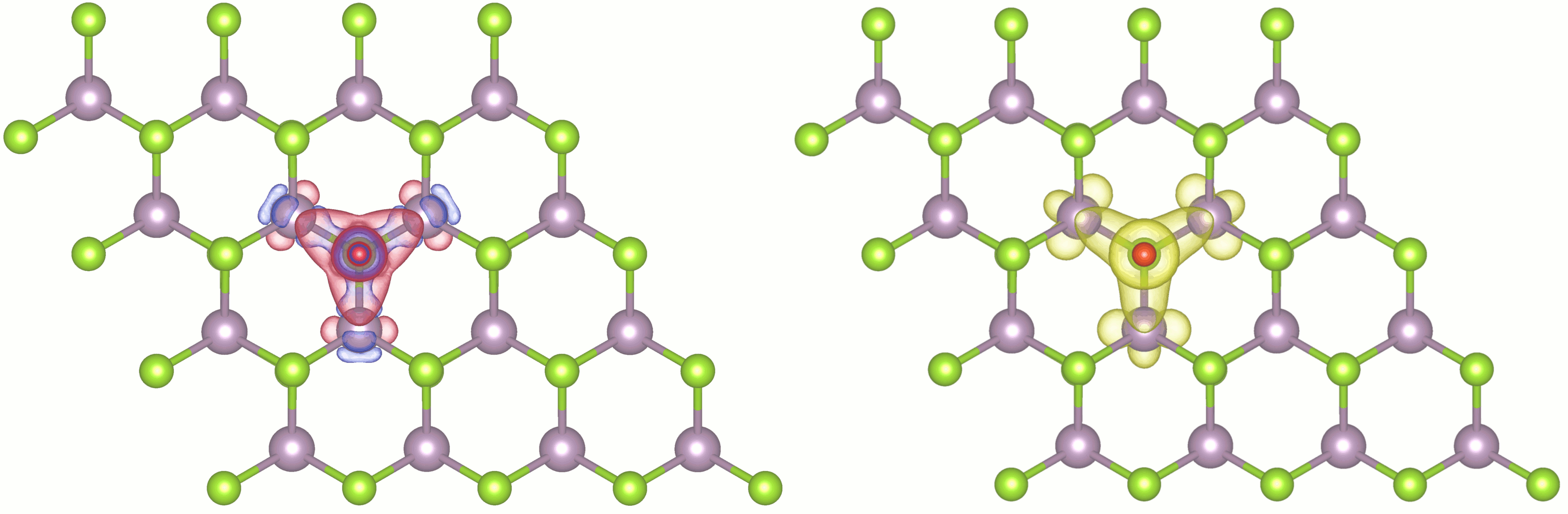}
\caption{PBE-D3 isosurfaces of the induced electron density $\Delta n$ and spin-magnetization density $\Delta m$ for CO adsorbed at the vacancy on MoSe$_2$-vSe. Red (Blue) isosurfaces indicate defect (excess) of electrons and yellow (blue) isosurfaces spin majority (minority) excess. Atoms colored as in Fig.~\ref{fig:bands_CO}. The isovalue is  6$\times 10^{-3}$ electrons/{\AA}$^{-3}$ in all cases.}
	\label{fig:charge_trans_CO}
\end{figure}

\begin{figure}
	\centering
	\includegraphics[width=0.95\linewidth]{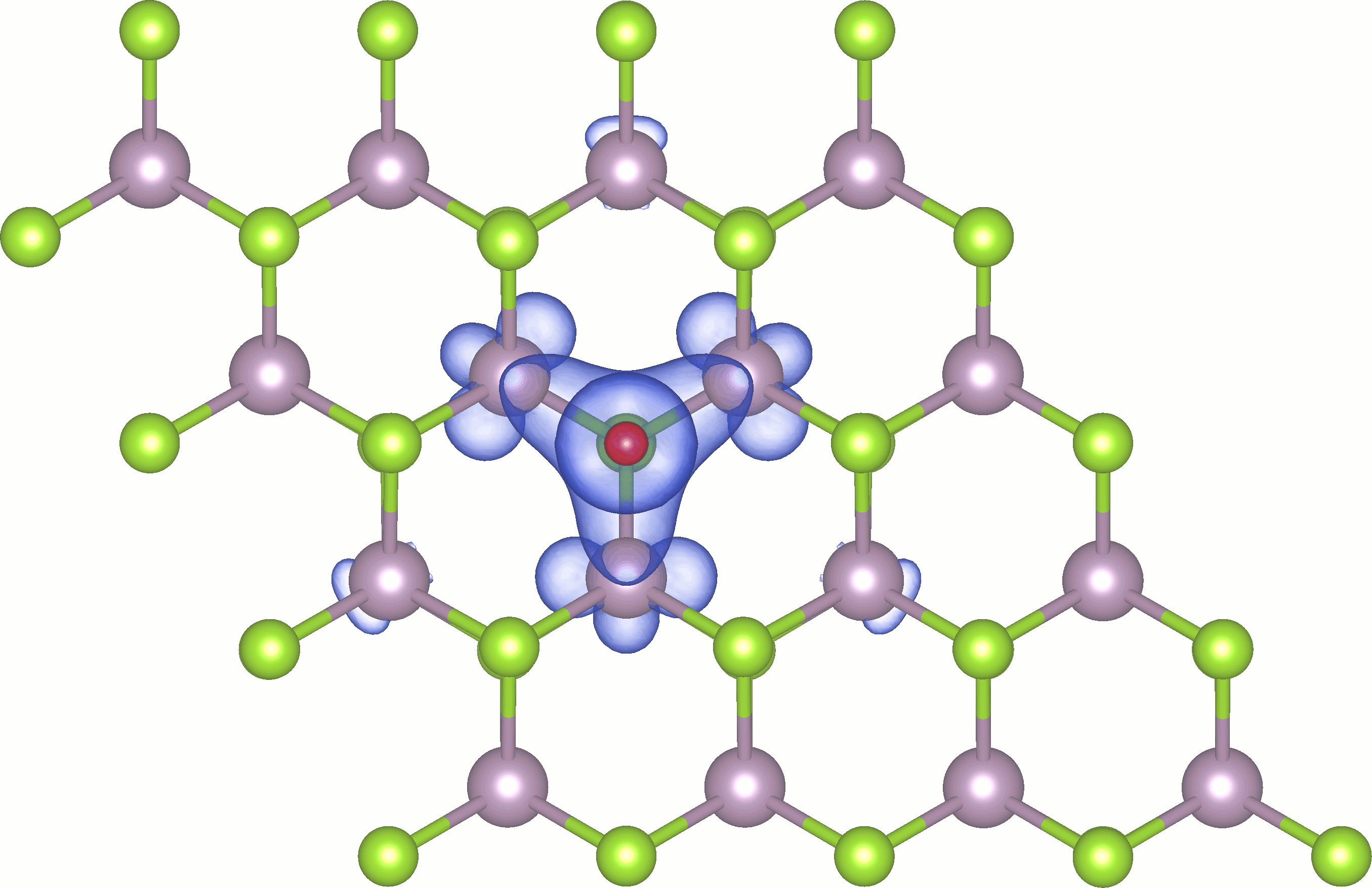}
	\caption{PBE-D3 partial electron density corresponding to the two CO- and Mo-contributed quasi-degenerated spin-majority bands that appear below the Fermi level upon CO adsorption at the Se vacancy.  Atoms colored as in Fig.~\ref{fig:bands_CO}. The isosurface value is 6$\times 10^{-3}$ electrons/{\AA}$^{-3}$. }
	\label{fig:parcharge_CO}
\end{figure}

The adsorption of CO in the MoSe$_2$ monolayer has been previously studied in Ref.~\cite{Ai2019} with PBE-D2. The reported adsorption energy is $E_a\sim$~$-$0.100~eV with the CO located at $Z\sim$~3~{\AA} in between the  $\mathrm{T_b}$ and $\mathrm{T_h}$ positions, and with a Bader charge transfer compatible with the one obtained in this work. Additionally, theoretical studies for CO adsorption in the MoS$_2$ and WS$_2$ monolayers report adsorption energies of the same order~\cite{Bui_2015,Zhou2015,Zhao2014}.

As also obtained for H$_2$ and O$_2$, CO adsorption at a Se vacancy is characterized by a much larger binding energy, although there is a large discrepancy between the revPBE-vdW ($-0.871$~eV) and PBE-D3 ($-1.317$~eV) values in this case. However, both calculations agree on the structural properties of the adsorption configuration, namely, $Z\sim0.2$~{\AA}, d(C-O)~$\sim 1.25$~{\AA}, and perpendicular orientation [see the $\mathrm{T_{vSe}}$ site in Table~\ref{table:CO_MoSe2} and Fig.~\ref{fig:bands_CO}(b)]. The normal mode analysis reveals that CO adsorption is stable at low temperatures, being all the modes $\hbar\omega>$~20~meV. In particular, for the mode related to desorption (i.e., translation perpendicular to the plane), $\hbar\omega\sim 35$~meV. Remarkably, upon adsorption of CO at the Se vacancy, the system acquires a spin-magnetic moment of 2~$\mu_B$. A Bader charge analysis reveals that the charge transfer that can be attributed to CO is  $\Delta\mathcal{Q}_\mathrm{B}$ = 0.680~$e$. It is one order of magnitude larger than the Bader charge transfer to CO adsorbed in MoSe$_2$, but insufficient to explain the system spin-magnetic moment (2 $\mu_B$). Figure~\ref{fig:charge_trans_CO} 
provides some insight into this point. The induced electron density shows that CO binds to the three closest Mo atoms below. Hence, although part of the spin-magnetization is located on the adsorbate, 
it also involves these three Mo atoms, as clearly shown in the figure. The band structure and DOS shown in Fig.~\ref{fig:bands_CO}(b) allow us to understand better the spin-magnetization. The MoSe$_2$-vSe in-gap states disappear, giving rise to two nearly flat and degenerate bands per spin component close to the Fermi level that are mainly contributed by $p$-like states of CO and $d$-like states of the three closest Mo. Figure~\ref{fig:parcharge_CO} shows the partial charge density associated with the two new bands appearing below the Fermi level to illustrate the shape and spatial distribution of these newly formed states. These bands, being partially occupied, are at the origin of the finite magnetic moment and transform the system into a $p$-type semiconductor. Besides, the minimum energy excitation is a spin excitation.

The band structure of CO adsorbed at a S vacancy on MoS$_2$ has been studied with DFT-D2~\cite{Li2016}. As obtained here for MoSe$_2$-vSe, the in-gap states vanish while new flat bands appear just below and above the Fermi level. In contrast to MoSe$_2$-vSe, the new bands are non-degenerated and, as a result, no spin-polarization appears. Similarly a non-magnetic ground state has been predicted for CO adsorption at the chalcogen vacancy position of  WSe$_2$~\cite{Ma2017} using the optB86b-vdW functional.

\begin{table*}
\caption{Properties of the proposed configurations for dissociative adsorption of CO on MoSe$_2$ [$\mathrm{T_{b^\prime}}(\mathrm{C}) + \mathrm{T_{ah}}(\mathrm{O})$ and $\mathrm{T_b}(\mathrm{C})+\mathrm{T_a}(\mathrm{C})$] and on MoSe$_2$-vSe [$\mathrm{T_{b^\prime}} (\mathrm{C}) + \mathrm{T_{vSe}}(\mathrm{O})$ , $\mathrm{T_b} (\mathrm{C})+ \mathrm{T_{vSe}}(\mathrm{O})$,  and $\mathrm{T_{vSe}}(\mathrm{C})+ \mathrm{T_a}(\mathrm{O}) $] (see text for details): Dissociation energies $E_d$ and $E_d^\infty$, C-O internuclear distance d(C-O), height of each adsorbate from the surface $z_\mathrm{C}$ and $z_\mathrm{O}$, and system spin-magnetic moment $|\vec{M}|$.}
\label{table:dissCO_MoSe2}
\begin{ruledtabular}
\begin{tabular}{ l ccccccc c ccccccc}
	  &\multicolumn{7}{c}{\textbf{revPBE-vdW}}& & \multicolumn{7}{c}{\textbf{PBE-D3}}\\\cline{2-8}\cline{10-16}
C+O$\rightarrow$	&	  $\mathrm{T_{b^\prime}} + \mathrm{T_{ah}}$& $\mathrm{T_b} +\mathrm{T_a}$  
	& & $\mathrm{T_{b^\prime}} + \mathrm{T_{vSe}}$ & $\mathrm{T_b} + \mathrm{T_{vSe}}$ 
	&  &   $\mathrm{T_{vSe}}+ \mathrm{T_a} $
	& &	  $\mathrm{T_{b^\prime}} + \mathrm{T_{ah}}$& $\mathrm{T_b} +\mathrm{T_a}$  
	& & $\mathrm{T_{b^\prime}} + \mathrm{T_{vSe}}$ & $\mathrm{T_b} + \mathrm{T_{vSe}}$ 
	&  &   $\mathrm{T_{vSe}}+ \mathrm{T_a} $ \\\cline{2-3} \cline{5-6}\cline{8-8}\cline{10-11}\cline{13-14}\cline{16-16}
	$E_d$(eV)	     & 6.700 & 6.589 & &2.066& 2.115  & & 1.620 & & 5.700 & 6.087 & & 1.671 & 1.667 & &1.263\\
	$E_d^\infty$(eV) &6.607    &  6.607 & & 2.474 &  2.474 & &  1.635   & & 6.114 & 6.114 &  &1.730 &1.730 & & 1.290\\
	d(C-O)(\AA)					& 3.26& 4.23 & & 2.48 & 3.93 & & 4.18 & &  3.11& 4.06 & &  2.44 &3.83& & 4.06\\
	$z_\mathrm{C}$(\AA) 			& 0.31& 0.04 & &  0.52&0.11 & &$-$0.90& &0.25&0.03 & & 0.49  &0.02 &  &$-$0.89\\
	$z_\mathrm{O}$(\AA) 			& 1.92&1.60  & & $-$0.73 &$-$0.66 & &1.60& &1.89&1.53 & & $-$0.76 &$-$0.68& &1.52\\
	$|\vec{M}|$($\mu_B$)	 	& 0 & 2 	 & & 0&  2 & & 0& & 0 & 2&  &  0&2 &  &0
\end{tabular}
\end{ruledtabular}
\end{table*} 
The adsorption energies for atomic O and C are written in Appendix~\ref{app:at_ener} (see Table~\ref{table:at_ener}). The most favorable positions on pristine MoSe$_2$ are $\mathrm{T_a}$ and $\mathrm{T_b}$ for O and C, respectively. Therefore, as dissociative adsorption configuration of CO on the pristine monolayer we consider that each atom adsorbs at its most energetically favorable position. Since the separation between adjacent $\mathrm{T_b}$ and $\mathrm{T_a}$ sites is smaller than one lattice constant, we account for the two following configurations. In the first one, denoted $\mathrm{T_{b^\prime}(C)} + \mathrm{T_a(O)}$, C and O adsorb each atop first nearest Mo and Se neghbors, respectively. In the second one, $\mathrm{T_b(C)} + \mathrm{T_a(O)}$, C and O are adsorbed atop second nearest Mo and Se neighbors. Since the O in the first configuration ends at the $\mathrm{T_{ah}(O)}$ position after relaxation, in the following we will refer to this configuration as $\mathrm{T_{b^\prime}(C)} + \mathrm{T_{ah}(O)}$. All these results are summarized in Table~\ref{table:dissCO_MoSe2}. Even if the dissociation energy of CO in the pristine MoSe$_2$ surface is roughly a factor two smaller than the gas-phase value, the process remains highly endothermic. In particular, revPBE-vdW predicts $E_d=$ 6.589~eV for $\mathrm{T_b(C)} + \mathrm{T_a(O)}$ as the lowest dissociation energy, while $\mathrm{T_{b^\prime}(C)} + \mathrm{T_{ah}(O)}$ with $E_d=5.700$~eV is the energetically favored dissociation configuration according to PBE-D3. There are differences between $\mathrm{T_b^\prime(C)} + \mathrm{T_{ah}(O)}$ and $\mathrm{T_b(C)} + \mathrm{T_a(O)}$ that not only refer to the energetics, but also to the C and O adsorption height, the induced surface distortions, and remarkably, the spin-magnetization that changes from zero to 2$\mu_\mathrm{B}$ as the C-O distance increases. According to revPBE-vdW and PBE-D3, the dissociation energy, the overall adsorption structure, and the total magnetic moment for $\mathrm{T_b(C)} + \mathrm{T_a(O)}$ (i.e., C and O adsorbed at second nearest Mo-Se neighbors) are already very similar to the results obtained when considering the atoms adsorbed at infinitely separated sites (compare the $\mathrm{T_b(C)} + \mathrm{T_a(O)}$ properties to $E_d^\infty$ and to the C and O atomic adsorption properties summarized in Table~\ref{table:at_ener} in Appendix~\ref{app:at_ener}).
\begin{figure*}
	\includegraphics[width=1.0\linewidth]{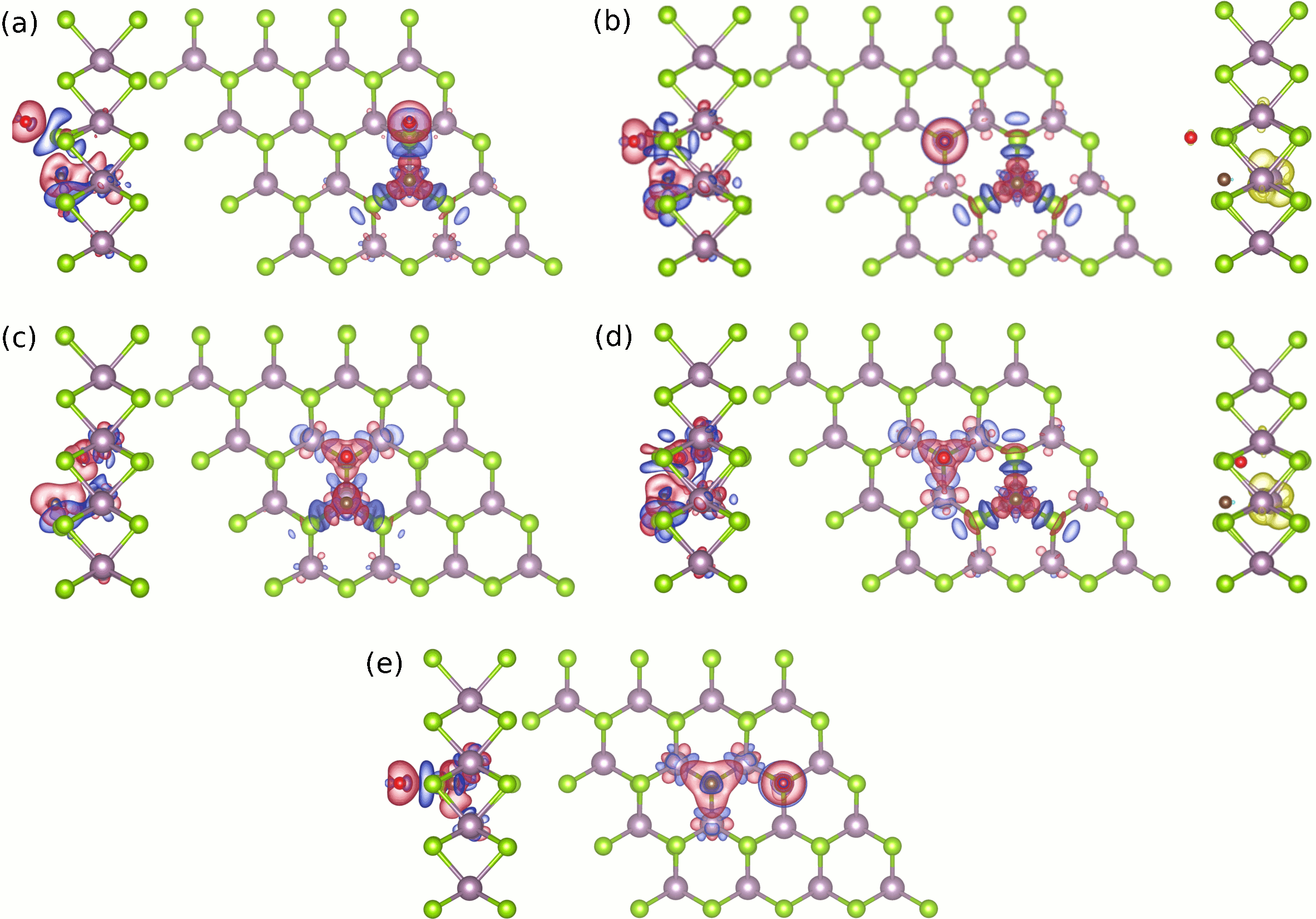}
\caption{PBE-D3 isosurfaces (top and side views) of the induced electron density $\Delta n$ for CO dissociated on MoSe$_2$ at (a)  $\mathrm{T_{b^\prime}} (\mathrm{C})+ \mathrm{T_{ah}}(\mathrm{O})$, (b)  $\mathrm{T_b}(\mathrm{C}) + \mathrm{T_{a}}(\mathrm{O})$, and on MoSe$_2$-vSe at (c) $\mathrm{T_{b^\prime}}(\mathrm{C}) + \mathrm{T_{vSe}}(\mathrm{O})$, (d)  $\mathrm{T_b}(\mathrm{C}) + \mathrm{T_{vSe}}(\mathrm{O})$, and (e) $\mathrm{T_vSe}(\mathrm{C}) + \mathrm{T_{a}}(\mathrm{O})$. In (b) and (d) also the spin-magnetization density $\Delta m$ (side views) is shown. Red (Blue) isosurfaces indicate defect (excess) of electrons and yellow (blue) isosurfaces spin majority (minority) excess. Atoms colored as in Fig.~\ref{fig:bands_CO}. An isovalue of 6$\times 10^{-3}$ electrons/{\AA}$^{-3}$ is used in all cases.}
	\label{fig:charge_dis_CO}
\end{figure*}

The analysis of the charge transfer distribution allows us to understand better the differences  between the  $\mathrm{T_{b^\prime}(C)} + \mathrm{T_{ah}(O)}$  and $\mathrm{T_b(C)} + \mathrm{T_a(O)}$ configurations, including the very different spin magnetization. The Bader charge analysis assigns a rather similar charge to C and also to O in the two situations. In $\mathrm{T_{b^\prime}(C)} + \mathrm{T_{ah}(O)}$, the C and O Bader charge transfer are 0.75~$e$  and 0.88~$e$, respectively, whereas 0.82~$e$ and O 0.94~$e$ are the corresponding values when the atoms adsorb at second nearest Mo-Se neighbors. The induced electron density plotted in Figs.~\ref{fig:charge_dis_CO}(a) and (b) clarifies that the differences arise from the different surface atoms involved in the bindings. In $\mathrm{T_{b^\prime}(C)} + \mathrm{T_{ah}(O)}$, the strong bond between the O atom and its nearest Se, which is also close to C, breaks the three-fold symmetry of the $\mathrm{T_{b^\prime}(C)}$ site, restricting the binding of the C atom to the two other Se atoms [see the side view in Fig.~\ref{fig:charge_dis_CO}(a)]. In contrast, since O binds to a different Se in $\mathrm{T_b(C)} + \mathrm{T_a(O)}$, the C atom forms strong bonds with its three nearest Se. The charge rearrangement induced by these new bonds also extends to the Mo beneath. The corresponding spin magnetization density of Fig.~\ref{fig:charge_dis_CO} shows that the finite spin-magnetic moment is in fact localized at this Mo atom. We have verified that this is exactly the charge and spin rearrangement that a single C atom induces when adsorbing on the pristine surface.

Regarding the dissociative adsorption of CO on MoSe$_2$-vSe, Table~\ref{table:at_ener} in Appendix~\ref{app:at_ener} shows that the adsorption energies of the two atomic species at the Se vacancy are similar and highly exothermic ($E_a\sim - 7.6$~eV with PBE-D3, for instance). Therefore, we consider the following dissociatively adsorbed configurations:  C is at the vacancy and O at its also stable $\mathrm{T_a}$ site [$\mathrm{T_{vSe}(C)} + \mathrm{T_a(O)}$], O is at the vacancy and C is either at the first [$\mathrm{T_{b^\prime} (C)} + \mathrm{T_{vSe}(O)}$] or the second [$\mathrm{T_b (C)} + \mathrm{T_{vSe}(O)}$] nearest $\mathrm{T_b}$ site from the vacancy.  
The properties of the optimized configurations are summarized in Table~\ref{table:dissCO_MoSe2}. Although the dissociation process close to the vacancy is still endothermic, there is a significant reduction in $E_d$ that now varies within 1--2~eV, depending on the configuration and calculation method. Among the studied configurations, $\mathrm{T_{vSe}(C)} + \mathrm{T_a(O)}$ with a revPBE-vdW (PBE-D3) dissociation energy of 1.620 (1.263)~eV is the less endothermic. The corresponding $E_d^{\infty}$ value differs in less than 15 (30)~meV, suggesting that the interaction between the adsorbed atoms (and also between the perturbations induced locally) is minor. The $\mathrm{T_{vSe}(C)} + \mathrm{T_a(O)}$ induced electron density is shown in Fig.~\ref{fig:charge_dis_CO}(e). Clearly, the local perturbation created by the adsorbed O is similar to the one observed for $\mathrm{T_{b} (C)} + \mathrm{T_a (O)}$ in the pristine MoSe$_2$ surface [Fig.~\ref{fig:charge_dis_CO}(b)]. This similarity is in accordance with the aforementioned minor interaction between the adsorbates. The C and O Bader charge transfer are 0.86~$e$  and 0.79~$e$ in this configuration.

As Table~\ref{table:dissCO_MoSe2} shows, when O is adsorbed at the Se vacancy, $E_d$ does not significantly depend on whether C is adsorbed at $\mathrm{T_{b^\prime}}$ or $\mathrm{T_{b}}$. However, the spin magnetization does. The comparative analysis of the induced electron density between both configurations shows that the situation is similar to what is observed for $\mathrm{T_b (C)} + \mathrm{T_a(O)}$ and $\mathrm{T_b^{\prime}(C)} + \mathrm{T_a(O)}$ in MoSe$_2$ [cf., Figs.~\ref{fig:charge_dis_CO}(c) and (d)]. When the adsorbed C and O are well separated, the C atom binds to the three nearest Se, causing a strong spin-polarized charge rearrangement around the Mo below. In this situation the C and O Bader charge transfer are 0.75~$e$  and 1.01~$e$, respectively. 

It is important to remark that in the $\mathrm{T_{b^\prime}(C)} + \mathrm{T_{vSe}(O)}$ configuration the distance between the two ions is only d(C-O)~$\sim$~2.4~{\AA}. Therefore, this configuration might be considered as an intermediate step for dissociation. The Bader charge transfer to C and O are reduced being 0.62~$e$ and 0.99~$e$ respectively. Remarkably, if one considers that two Se vacancies are present at neighbor positions in the surface, the process becomes exothermic. The revPBE-vdW (PBE-D3)  dissociation energy becomes $E_d~= -2.661$ ($-$3.036)~eV. This result is close to what one obtains considering that the two Se vacancies are isolated $E_d^\infty\sim$~$-$2.498 ($-$3.094)~eV.

\subsection{NO adsorption and dissociation} 
\label{subsect:NO}

\begin{figure*}[tb!]
	\centering
	\includegraphics[width=1.0\linewidth]{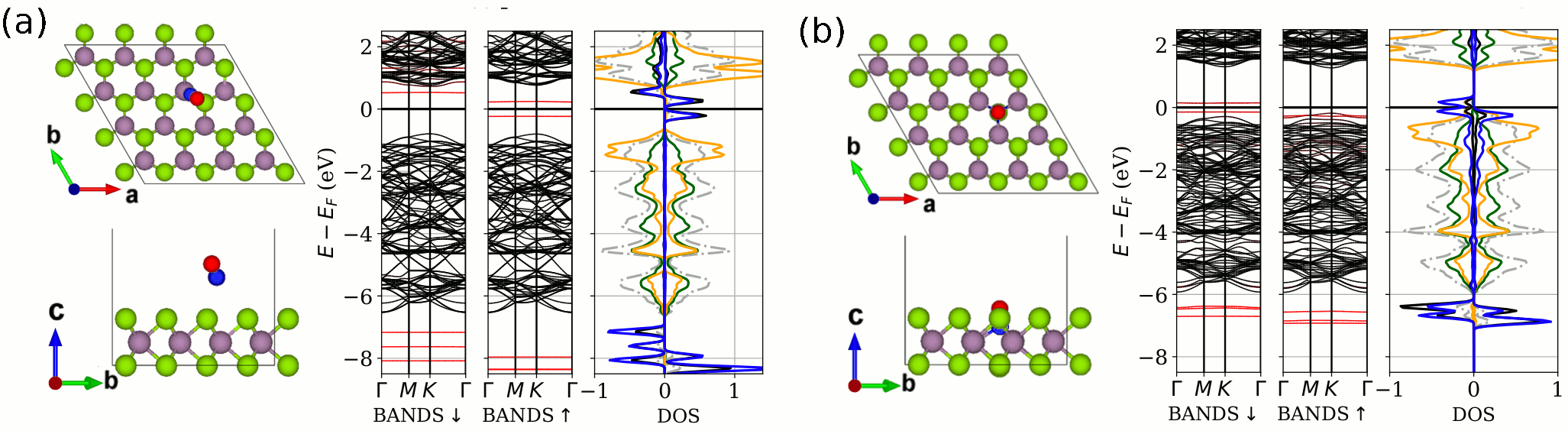}
\caption{PBE-D3 optimized structure (top and side views), band structure, and DOS for NO adsorbed at (a) T$_\mathrm{ab}$ on  MoSe$_2$ and (b) the vacancy on MoSe$_2$-vSe. Se, Mo, N, and O atoms colored in green, purple, blue, and red, respectively. Bands in red correspond to projection onto the adsorbate states. Total DOS in grey, Mo-projected $d$-states in orange, Se-projected, N-projected, and O-projected $p$-states in green, black, and blue, respectively. Spin-up (spin-down) bands and positive (negative) DOS refer to the spin majority (minority) components. For clarity, all the density of states distributions are divided by the number of ions that involve, except for N and O that are divided by three.}
	\label{fig:bands_NO}
\end{figure*}

\begin{table}[tb!]
\caption{Properties of the optimized configurations for NO adsorption on the pristine MoSe$_2$ surface ($\mathrm{T_a}$ to $\mathrm{T_{ah}}$ sites) and on MoSe$_2$-vSe ($\mathrm{T_{vSe}}$): Adsorption energy $E_a$, bond length d(N-O), adsorption height $Z$, molecular orientation ($\vartheta$, $\varphi$), and spin-magnetic moment $|\vec{M}|$.}
\label{table:NO_MoSe2}
\begin{ruledtabular}
\begin{tabular}{ l c c c c c c c}
	\multicolumn{8}{c}{\textbf{revPBE-vdW}}\\
	\hline
	& $\mathrm{T_a}$& $\mathrm{T_b}$& $\mathrm{T_h}$& $\mathrm{T_{ab}}$& $\mathrm{T_{ah}}$& &$\mathrm{T_{vSe}}$\\  \cline{2-6}\cline{8-8}
	$E_a$(eV)			& $-$0.201& $-$0.212& $-$0.216& $-$0.220& $-$0.216& &$-$2.471\\
	d(N-O)({\AA})		& 1.163	& 1.161	& 1.161	& 1.162	& 1.163 & & 1.276 \\
	$Z$({\AA})					& 3.84	& 3.66	& 3.64	& 3.51	& 3.55	& & 0.09\\
	$\vartheta$  & 51$^\circ$ & 36$^\circ$ & 56$^\circ$ & 36$^\circ$ & 43$^\circ$ & & 3$^\circ$\\
	$\varphi$  &$-$48$^\circ$ &$-$86$^\circ$ &$-$87$^\circ$ &$-$60$^\circ$ & 5$^\circ$ & &$-$87$^\circ$\\
	$|\vec{M}|$($\mu_B$)				& 1		& 1		&	1	&	1	&	1	& &	1\\
	\multicolumn{7}{c}{\vspace{0.0cm}}\\
	\multicolumn{8}{c}{\textbf{PBE-D3}}\\	
	\hline
	& $\mathrm{T_a}$& $\mathrm{T_b}$& $\mathrm{T_h}$& $\mathrm{T_{ab}}$& $\mathrm{T_{ah}}$& &$\mathrm{T_{vSe}}$\\  \cline{2-6}\cline{8-8}
	$E_a$(eV)			& $-$0.185& $-$0.176 &  $-$0.195 &  $-$0.200 &  $-$0.204 &  &$-$2.778\\
	d(N-O)({\AA})		& 1.159	& 1.157	& 1.158	& 1.159 & 1.159	& & 1.265 \\
	$Z$({\AA})			& 3.38	& 3.19	& 3.14	& 3.16	& 3.12& & 0.05	\\
	$\vartheta$  & 51$^\circ$ & 20$^\circ$ & 54$^\circ$ & 30$^\circ$ & 45$^\circ$ & & 4$^\circ$\\
	$\varphi$  &$-$47$^\circ$ &$-$73$^\circ$ & 46$^\circ$ & $-$66$^\circ$ & 87$^\circ$ & & 86$^\circ$\\
	$|\vec{M}|$($\mu_B$)				& 1		& 1		&	1	&	1	&	1	& &	1\\
\end{tabular}
\end{ruledtabular}
\end{table} 

The revPBE-vdW (PBE-D3) prediction for the NO$_{\mathrm{(gas)}}$ bond length is 1.162 (1.159)~{\AA}, in good agreement with the experimental value (1.154~{\AA})~\cite{NIST}. In contrast, the revPBE-vdW (PBE-D3) dissociation energy of gas-phase NO, $E_d= 7.586$ (7.748)~eV, overestimates the experimental $E_d=6.665$~eV  that we calculate by adding the NO$_\mathrm{(gas)}$ ZPE of 117~meV~\cite{Irikura2007} to the experimental value of 6.548~eV \cite{Benson1965}.

Similarly to what we obtained for the previous diatomic molecules, the structural relaxation suggests that NO may adsorb on any of the five sites that were studied in MoSe$_2$. The results are  summarized in Table~\ref{table:NO_MoSe2}. All the relaxation calculations start with the NO molecular axis perpendicular to the surface, being the N atom the closest to the surface. According to revPBE-vdW, the energetically preferred adsorption site is $\mathrm{T_{ah}}$ with $E_a=-0.220$~eV. In contrast, PBE-D3 predicts that the most energetically favorable position is in between a Se-Mo bond ($\mathrm{T_{ab}}$), being $E_a=-0.204$~eV. Nonetheless, the difference between the $\mathrm{T_{ab}}$ and $\mathrm{T_{ah}}$ adsorption energies is only 4~meV in each case. This small difference is possibly beyond the numerical precision and we can consider that the two positions are degenerated. Actually, the adsorption energies of the five sites differ in less than 20~meV, suggesting a rather planar potential energy surface.  In this respect, a normal mode analysis for $\mathrm{T_{ab}}$ yields soft modes for translation on the surface and rotation around the surface normal ($\hbar\omega\le$~8meV). Furthermore, the molecule is weakly bounded along the surface normal [$\hbar\omega\sim$~15 (12)~meV with revPBE-vdW (PBE-D3)]. With $\hbar\omega\sim$~20~meV, the mode associated to NO tilting is slightly higher. Altogether, these results and the minor Bader charge transfer ($\Delta\mathcal{Q}_\mathrm{B} =$ 0.05~$e$) imply that NO is weakly physisorbed on MoSe$_2$.

Figure~\ref{fig:bands_NO}(a) shows the optimized structure for NO adsorbed on MoSe$_2$ at $\mathrm{T_{ab}}$ together with the corresponding band structure and DOS. The aforementioned weak interaction agrees with the lack of dispersion in the NO-dominated bands (red bands and corresponding sharp peaks in the $p$-like contribution of the N- and O-projected DOS) and the fact that these bands do not overlap with the MoSe$_2$ ones. 
We note that  the adsorption of NO in the MoSe$_2$ monolayer has been studied in Refs.~\cite{Panigrahi2019,Ai2019} using PBE-D2. In general, the reported adsorption properties are in good agreement with our results. The exception is the adsorption energy, which is smaller for PBE-D2, as also found in adsorption studies performed on other TMD monolayers by Zhao \textit{et al.}, using different functionals~\cite{Zhao2014}. The revPBE-vdW and PBE-D3 adsorption energies of NO on the MoS$_2$ monolayer calculated by these authors are similar to the ones that we obtain for MoSe$_2$. 

\begin{figure}[tb!]
	\centering
	\includegraphics[width=1.0\linewidth]{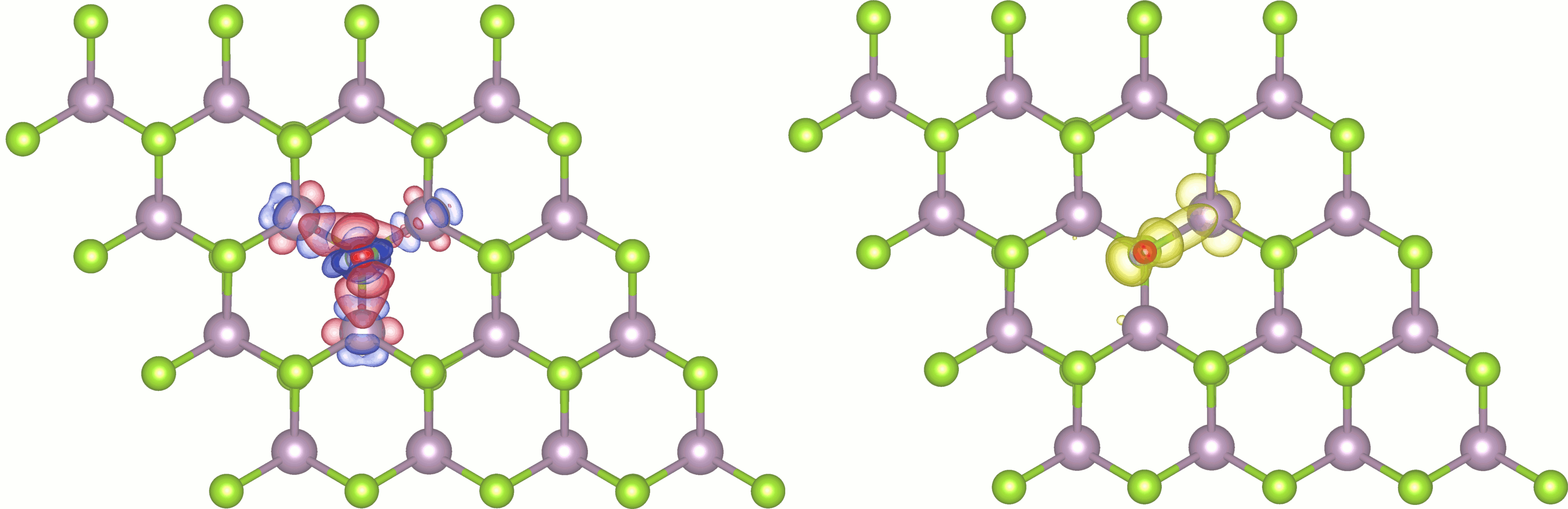}
\caption{PBE-D3 isosurfaces of the induced electron density $\Delta n$ and spin-magnetization density $\Delta m$ for NO adsorbed (a) at $\mathrm{T_{ab}}$ on MoSe$_2$ and (b) at the Se vacancy on MoSe$_2$-vSe. Red (Blue) isosurfaces indicate defect (excess) of electrons and yellow (blue) isosurfaces spin majority (minority) excess. Atoms colored as in Fig.~\ref{fig:bands_NO}. An isovalue of 6$\times 10^{-3}$ electrons/{\AA}$^{-3}$ is used in all cases.}
	\label{fig:charge_trans_NO}
\end{figure}

Also in Table~\ref{table:NO_MoSe2} we report the results for NO adsorption on MoSe$_2$-vSe ($\mathrm{T_{vSe}}$). The molecule is slightly elongated and stays nearly perpendicular to the surface with the N atom well inside the vacancy, as can be observed in Fig.~\ref{fig:bands_NO}(b). The revPBE-vdW (PBE-D3) adsorption energy $E_a=-2.471$ ($-2.778$)~eV suggests that adsorption on the vacancy is quite stable. The normal mode analysis confirms the stability at low temperatures, since all the modes have energies, $\hbar\omega>$~20~meV. In particular, the mode related to desorption (translation perpendicular to the surface) is $\hbar\omega\sim$~40~meV. Noticeably, the spin magnetic moment is $1\mu_\mathrm{B}$ as in the gas-phase and physisorbed NO cases. At first sight, this result surprises because the Bader charge analysis assigns 0.74~$e$ to NO at this position. The induced electron density and spin-magnetization density in Fig.~\ref{fig:charge_trans_NO} show that NO binds to the three nearest Mo and, as a result, the spin moment is not longer localized on NO, but it also involves one Mo.

\begin{figure}[tb]
\begin{flushleft}
	\includegraphics[width=0.95\linewidth]{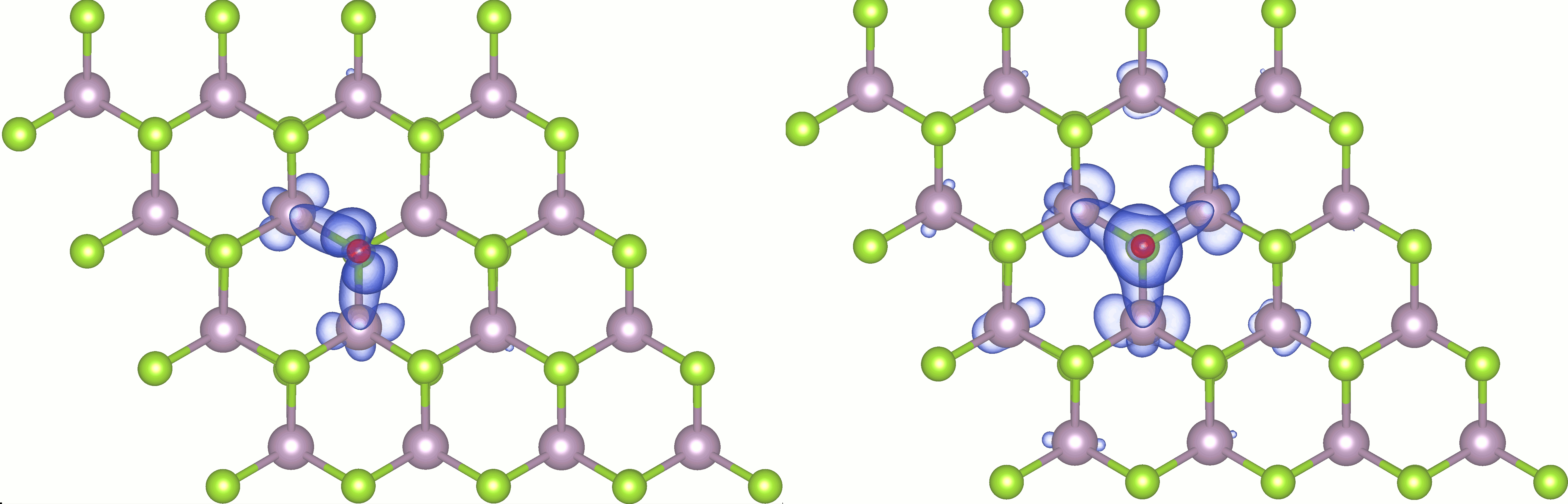}
\end{flushleft}
	\caption{PBE-D3 partial electron density corresponding to the NO- and Mo-contributed bands that appear below the Fermi level upon NO adsorption at the Se vacancy. Left (right) plot corresponds to the minority (majority) spin component. Atoms colored as in Fig.~\ref{fig:bands_NO}. The isosurface value is 6$\times 10^{-3}$ electrons/{\AA}$^{-3}$. }
	\label{fig:parchar_NOvac}
\end{figure}

The band structure and DOS represented in Fig.~\ref{fig:bands_NO}(b) clarify the formation and properties of these bonds. In common to the case of CO adsorbed at the vacancy, the MoSe$_2$-vSe in-gap states disappear, while two nearly flat and degenerate new states appear. The atom-projected DOS shows that these states involve the $p$-like orbitals of NO and the $d$-like orbitals of the Mo atoms beneath. As for CO chemisorption, the system acquires $p$-type semiconductor character. The main difference to the CO case is that not only the spin-majority bands, but also one of the spin-minority are occupied, thus a spin excitation is not the lowest energy one. The partial electron density of the quasi-degenerate spin-up bands and the occupied spin-down band are represented in Fig.~\ref{fig:parchar_NOvac}, explaining the spin-magnetization density of the system [see Fig.~\ref{fig:charge_trans_NO}(b)]. 
Moreover, the highest occupied state that can be mostly attributed to the MoSe$_2$ surface exhibits its maximum at the $M$-high-symmetry point. This is different from the valence band maximum of the isolated surface that occurs at the high-symmetry $K$-point.

Recent DFT studies that account for vdW corrections in the adsorption of NO at the chalcogen vacancy of monolayer MoS$_2$~\cite{Li2016,Ma2016,Li2018} and WSe$_2$~\cite{Ma2017} predict adsorption energies in the order of the ones obtained here. Interestingly, the overall appearance of the MoS$_2$ band structure around the Fermi level when NO adsorbs at the S vacancy of MoS$_2$~\cite{Li2016} resembles that of Fig.~\ref{fig:bands_NO}(b).
\begin{table}[tb!]
    \caption{Properties of the proposed configurations for dissociate adsorption of NO on MoSe$_2$ [$\mathrm{T_a}(\mathrm{N}) +\mathrm{T_a}(\mathrm{O})$] and on MoSe$_2$-vSe [$\mathrm{T_a}(\mathrm{N}) + \mathrm{T_{vSe}}(\mathrm{O})$ and    $\mathrm{T_{vSe}}(\mathrm{N})+ \mathrm{T_a}(\mathrm{O})$] (see text for details): Dissociation energies $E_d$ and $E_d^\infty$, N-O internuclear distance d(N-O), height of each adsorbate from the surface $z_\mathrm{N}$ and $z_\mathrm{O}$, and system spin-magnetic moment $|\vec{M}|$.}
\label{table:dissNO_MoSe2}
\begin{ruledtabular}
\begin{tabular}{ l c  c   c   c}
	\multicolumn{5}{c}{\textbf{revPBE-vdW}}\\\hline
	N + O$\rightarrow$	&	& $\mathrm{T_a} +\mathrm{T_a}$ &  $\mathrm{T_a} + \mathrm{T_{vSe}}$ &    $\mathrm{T_{vSe}}+ \mathrm{T_a} $\\  \cline{3-3}\cline{4-5}
	$E_d$(eV)       & &3.784 & $-$0.711  &$-$1.927 \\
	$E_d^\infty$eV)  & &3.722 & $-$0.701 &$-$1.825 \\
	d(N-O)({\AA})	& &3.65 & 4.02 & 4.11\\
	$z_\mathrm{N}$({\AA})		& &1.77& 1.73 & $-$0.79\\
	$z_\mathrm{O}$({\AA})		& &1.62&  $-$0.67 & 1.60\\
	$|\vec{M}|$($\mu_B$)	 	& & 1 & 1 & 0 \\
	\hline
	\multicolumn{5}{c}{\vspace{0.0cm}}\\
	\multicolumn{5}{c}{\textbf{PBE-D3}}\\\hline
	N + O$\rightarrow$	&	& $\mathrm{T_a} +\mathrm{T_a}$ &  $\mathrm{T_a} + \mathrm{T_{vSe}}$ &    $\mathrm{T_{vSe}}+ \mathrm{T_a} $\\  \cline{3-3}\cline{4-5}
	$E_d$(eV)       & &3.626 & $-$0.838  &$-$2.043 \\
	$E_d^\infty$eV)  & &3.573 & $-$0.811 &$-$1.995 \\
	d(N-O)({\AA})	& & 3.56 & 4.02 & 4.02\\			
	$z_\mathrm{N}$({\AA})	& & 1.68& 1.64 & $-$0.81\\
	$z_\mathrm{O}$({\AA})	& & 1.54 & $-$0.70& 1.52\\
	$|\vec{M}|$($\mu_B$)	    & & 1& 1 & 0 
\end{tabular}
\end{ruledtabular}
\end{table}

\begin{figure}[tb]
\includegraphics[width=1.00\linewidth]{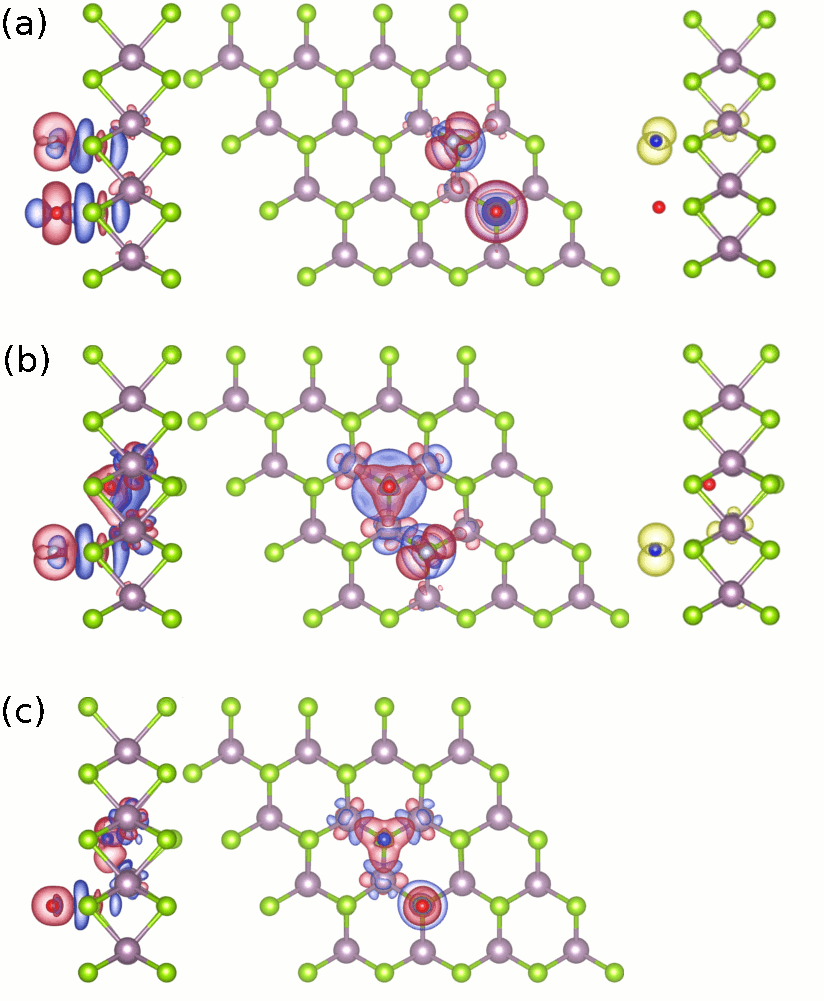}
\caption{PBE-D3 isosurfaces (top and side views) of the induced electron density $\Delta n$ for NO dissociated (a) on MoSe$_2$ at  $\mathrm{T_{a^\prime}} (\mathrm{N})+ \mathrm{T_{a}}(\mathrm{O})$, and on MoSe$_2$-vSe at (b) $\mathrm{T_{a}}(\mathrm{N}) + \mathrm{T_{vSe}}(\mathrm{O})$, (c)  $\mathrm{T_{vSe}}(\mathrm{N}) + \mathrm{T_{a}}(\mathrm{O})$. In (a) and (b) also the spin-magnetization density $\Delta m$ (side views) is shown. Red (Blue) isosurfaces indicate defect (excess) of electrons and yellow (blue) isosurfaces spin majority (minority) excess. Atoms colored as in Fig.~\ref{fig:bands_CO}. An isovalue of 6$\times 10^{-3}$ electrons/{\AA}$^{-3}$ is used in all cases.}
\label{fig:charge_dis_NO}
\end{figure}
Because $\mathrm{T_a}$ is the energetically preferred site for the atomic adsorption of N and O (see Appendix~\ref{app:at_ener}), we focus our study on the dissociative adsorption of NO on MoSe$_2$ to the case in which both N and O end adsorbed at different $\mathrm{T_a}$ positions. Our results are summarized in Table~\ref{table:dissNO_MoSe2}. The dissociation process at the MoSe$_2$ surface is still endothermic, being the revPBE-vdW (PBE-D3) dissociation energy $E_d=$~3.784 (3.626)~eV. The difference of around 60~meV between $E_d$ and $E_d^\infty$ values points to a very minor interaction between the adsorbed N and O. Note, for instance, that this difference is one order of magnitude smaller than the one we obtained for O$_2$ dissociation in the same $\mathrm{T_a(O)}+ \mathrm{T_a(O)}$ configuration. The induced electron density of the dissociated $\mathrm{T_a(N)} + \mathrm{T_{a}(O)}$ configuration in Fig.~\ref{fig:charge_dis_NO}(a) shows that each atom is strongly bounded to the Se below, being the Bader charge transfer for N and O, $\Delta\mathcal{Q}_\mathrm{B} =$~0.60 and 0.85~$e$, respectively. The spin-magnetization density in the same figure shows that the finite spin magnetic moment is located both around the N and one of the closest Mo. 

To study the dissociation of NO in the vicinity of a Se vacancy we consider two cases, namely, the O atom is adsorbed at the Se vacancy and the N atom at a $\mathrm{T_a}$ position separated by a lattice constant [$\mathrm{T_a(N)} + \mathrm{T_{vSe}(O)}$]  and vice verse [$\mathrm{T_{vSe}(N)}+\mathrm{T_a(O)}$].  In Table~\ref{table:dissNO_MoSe2}  we summarize our results. Interestingly, the process becomes exothermic in both cases, being the $\mathrm{T_{vSe}(N)}+\mathrm{T_a(O)}$ dissociated configuration the most exothermic with a revPBE-vdw (PBE) value, $E_d=-1.927 $ ($-2.043$)~eV. The corresponding induced density displayed in Fig.~\ref{fig:charge_dis_NO}(c) resembles that of the dissociative adsorption of CO at the $\mathrm{T_{vSe}(C)}+\mathrm{T_a(O)}$ configuration. The N atom, which adsorbs well inside the vacancy, binds to its three closest Mo, while the O atom does to the Se below. According to the Bader charge analysis N gains 1.08~$e$, slightly larger than the 0.86~$e$ obtained for C in the $\mathrm{T_{vSe}(C)}+\mathrm{T_a(O)}$ configuration. The charge transfer towards the O atom is the same as in the $\mathrm{T_{vSe}(N)}+\mathrm{T_a(O)}$ for the pristine surface. The zero spin magnetic moment in this configuration contrasts with the result obtained for  dissociation in the $\mathrm{T_a(N)} + \mathrm{T_{vSe}(O)}$ case. Figure~\ref{fig:charge_dis_NO}(b) shows clearly that the spin-magnetization is caused by the adsorption of N at the $\mathrm{T_a(N)}$ site, in accordance to the results obtained for dissociation in the pristine surface. Note also that the charge rearrangement induced in the vacancy is similar irrespective of being occupied by the N or the O atom. This also agrees with the similar Bader charge transfer assigned to O at the vacancy ($\Delta\mathcal{Q}_\mathrm{B} =$~1.00~$e$). On the other hand, the N atoms takes 0.69~$e$ of charge. 

Finally, as found for the rest of diatomic molecules, the dissociation energy will become highly exothermic if the N and O end adsorbed at two adjacent Se vacancies. The revPBE-vdW (PBE-D3) prediction in this case is $E_d = -6.167$ ($-6.461$)~eV. These values are similar to the ones obtained when one considers that the two Se vacancies are isolated from each other, $E_d^\infty = -6.388$$ ($-$6.533)$~eV.

\section{Conclusions} 
\label{sec:conclusions}

By means of density functional theory calculations we have studied the adsorption and dissociation of different diatomic molecules, H$_2$, O$_2$, CO, and NO, on the monolayer MoSe$_2$, including the case of having an isolated Se vacancy (MoSe$_2$-vSe). Our study was aimed to characterize not only the adsorption properties (i.e., adsorption geometry, energetics, and stability of the adsorbate), but also the effect that the adsorbate has on the monolayer electronic structure. 

In all the cases the MoSe$_2$ surface holds several active sites for molecular adsorption. The adsorption mechanism in the pristine surface is physisorption. In the presence of Se vacancies the physics gets richer as the diatomic molecules form strong chemical bonds with the MoSe$_2$-vSe sheet.  
Upon H$_2$ chemisorption at the Se vacancy, no noticeably changes in the MoSe$_2$-vSe electronic structure are observed. However, the situation is different when any of the other three molecules chemisorbs: The MoSe$_2$-vSe in-gap states disappear and new bonding states are formed near the valence band maximum. In the case of O$_2$, a single spin degenerated flat band appears that overlaps in energy with the valence band maximum of MoSe$_2$-vSe. As a consequence, the gap of the composed system is quite similar to that of the pristine MoSe$_2$ monolayer. 
In the case of CO and NO, two new and quasi-degenerated states form above the valence band maximum. These states, being partially occupied by 2 and 3 electrons, creates a spin magnetization of 2 and 1~$\mu_\mathrm{B}$ that is spatially localized in the region around the adsorbate. The remaining empty states that lie just above the Fermi level give a $p$-type semiconductor character to the system.  

Although the pristine MoSe$_2$ monolayer facilitates the dissociation process, it remains endothermic for the four molecules studied here. 
The dissociation energy is considerably reduced in the presence of a single Se vacancy. Still, only the dissociation of O$_2$ ($E_d=-3.947$~eV) and NO ($E_d= -2.043$~eV) becomes exothermic. Nonetheless, our calculations reveal that a larger concentration of Se vacancies favours the dissociation in all cases. In particular, dissociation at two adjacent Se vacancies is exothermic for the four molecules studied here.  
Finally, we note that adsorption of atomic H, C, and N, also induces a finite spin-magnetization in the system, but it is more localized than in the case of the molecules. 

\acknowledgments{ The authors acknowledge financial support by the Gobierno Vasco-UPV/EHU
Project No. IT1246-19 and the Spanish Ministerio de Ciencia e Innovación [Grant No. PID2019-107396GB-I00/AEI/10.13039/501100011033]. This research was conducted in the scope of the Transnational Common Laboratory (LTC) “QuantumChemPhys – Theoretical Chemistry and Physics at the Quantum Scale”. Computational resources were provided by the DIPC computing center.
}

\appendix
\section{Characterization of the MoSe$_2$ system}
\label{app:MoSe2_char}
\subsection{Bulk $2H$-MoSe$_2$}
\begin{table}[tb!]
\caption{Lattice parameters $a$ and $c$ and band-gap $E_g$ for the $2H$ phase of bulk MoSe$_2$. Together with the results of this work other theoretical and experimental results from the bibliography are listed for comparison.}
\label{table:bulk_parameters}
\begin{ruledtabular}
\begin{tabular}{ c c c c  }
 	  & $a $ (\AA)  & $c$ (\AA) & $E_g$ (eV)  \\ 
	\hline
	revPBE-vdW (this work)	                    & 3.386	& 13.890& 1.08 \\  
	PBE-D3 (this work)                     & 3.283 & 12.720 & 0.75   	\\
	PBE-D2 Refs.~\cite{Tongay2012,Kim2021}  & 3.317 & 13.032 & 0.84 \\
	GW Ref.~\cite{Kim2021}          & $-$ & $-$ & 1.15 \\
	LDA  Ref.~\cite{Boker2001} &3.299 &  12.938 & $-$ \\
	Experiment Ref.~\cite{Zemann1965}& 3.289 & 12.927 & $-$ \\
	Experiment Ref.~\cite{Kam1982}   & $-$ & $-$  &1.12\\
	Experiment Ref.~\cite{Baglio1982}   & $-$ & $-$  &1.22\\
	Experiment Ref.~\cite{Tonndorf2013} & $-$ & $-$ & 1.1\\
\end{tabular}
\end{ruledtabular}
\end{table}

The common bulk structure in group VI TMD semiconductors, including bulk MoSe$_2$, is the $2H$ phase. It consists of an hexagonal lattice formed by planes of Mo atoms sandwiched between two layers of Se atoms. The results of our revPBE-vdW and PBE-D3 calculations for the $2H$ phase of bulk MoSe$_2$ are compared in Table~\ref{table:bulk_parameters} to available experimental values. The experimental lattice parameters, $a= 3.289$~{\AA} and $c=12.927$~{\AA}~\cite{Zemann1965}, are better reproduced by PBE-D3 than by revPBE-vdW. In contrast, the latter provides a better value of the measured indirect band-gap ($E_g$ = 1.1-1.2~eV \cite{Kam1982,Baglio1982,Tonndorf2013}) than the former. For comparison, we also include in Table~\ref{table:bulk_parameters} the results obtained by other authors using the local density approximation (LDA)~\cite{Boker2001} and PBE-D2~\cite{Tongay2012,Kim2021}.

\subsection{$1H$-MoSe$_2$ monolayer}

\begin{figure*}[tb!]
	\includegraphics[width=1.00\linewidth]{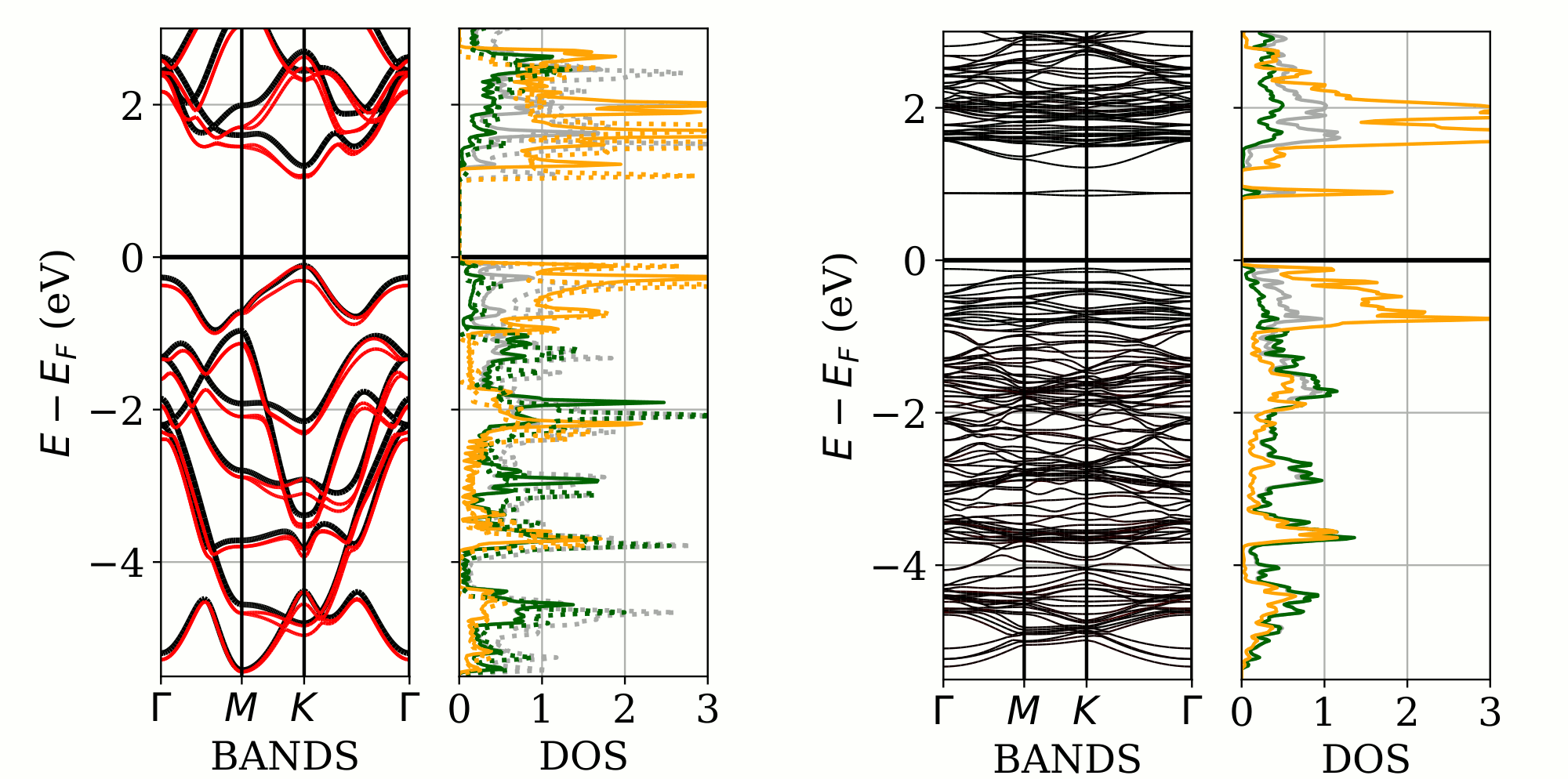}
	\caption{
		Left: band structure and DOS of the MoSe$_2$ monolayer,  evaluated in the primitive cell. Right: band structure and DOS of the  MoSe$_2$-vSe monolayer, evaluated in a $4\times4$ supercell. DOS color code: total  DOS in grey, Se-projected $p$-states in green, and Mo-projected $d$-states in orange. For MoSe$_2$ we also include results with spin-orbit coupling (red bands and dotted total and projected DOS). Each DOS is divided by the number of ions that involve to facilitate visualization. A gaussian smearing of 0.025~eV has been employed.}
	\label{fig:bands_vac}	
\end{figure*}

The MoSe$_2$ monolayer in the $1H$ phase \cite{Zhang2018} can be obtained, for example, by exfoliation of the $2H$ bulk phase. The atoms are organized in an hexagonal lattice with lattice constant $a$ equal to the bulk one~\cite{Ohuchi1991}. Although it is beyond the scope of this work, the phonon dispersion of the MoSe$_2$ monolayer has been previously studied in different DFT studies, demonstrating its dynamical stability~\cite{Yunguo2014,Ding2011,Mahrouche2021}.
In Table~\ref{table:2D_parameters} we compare our DFT results for the distances between the three atoms that are contained in the primitive cell to the DFT values obtained by other authors using different exchange-correlation functionals. 
In the last two columns of the same table, we show the results for the band-gap energy obtained from two different calculations, namely, $E_g$ corresponds to the value predicted by our standard DFT calculation that neglects the spin-orbit coupling (SOC) and  $E_g^\mathrm{SOC}$ is the value obtained when SOC is included using the {\sc vasp} implementation~\cite{Steiner2016}. The SOC introduces a splitting in the band structure at the high-symmetry $K$-point that reduces the band-gap in $\sim$100~meV. It is worth remarking that this is in good agreement with angle resolved photo-electron spectroscopy (ARPES) experiments showing that the MoSe$_2$ monolayer exhibits a direct band-gap  at the high symmetry $K$-point of $\sim$1.52~eV~\cite{Lu2014,Shaw2014}, with a splitting around the valence band maximum of $\sim$100--200~meV~\cite{Beal1972,Boker2001,Rasmussen2015}. This splitting is clearly observed in the PBE-D3 band structure of Fig~\ref{fig:bands_vac} (computed  in the primitive cell). The band structure evaluated with and without SOC (red and red black, respectively) are shown together to remark the differences that appear between the two calculations. In the same figure we include the DOS and its projection onto the $d$-states of Mo (orange) and $p$-states of Se (green). The minor effect that SOC causes in the adsorption and dissociation energies of the molecules studied in this work is analyzed in Appendix~\ref{sec:spin_orbit}.

\begin{table*}[tb!]
\caption{ DFT structural parameters for the MoSe$_2$ monolayer as obtained in this work and by other authors: lattice constant $a$, distance between nearest Mo and Se atoms d(Mo-Se), and distance between the two Se layers d(Se-Se). Last two columns: DFT band-gap energies evaluated without ($E_g$) and with (E$_g^\mathrm{SOC}$) SOC. Experimental band-gap values are written in a separate column.}
\begin{ruledtabular}
\begin{tabular}{ c c c c  c c}
   &$a $& d(Mo-Se) (\AA) & d(Se-Se) (\AA) &$E_g$ (eV) &  $E_g^\mathrm{SOC}$ (eV)\\  \hline
	revPBE-vdW (this work)                      & 3.386 &  2.579 & 3.363  & 1.30 & 1.17\\
	PBE-D3 (this work)                     & 3.282 & 2.531 & 3.357 & 1.54 & 1.45\\
	PBE-D2 Refs.~\cite{Tongay2012,Kim2021} & 3.320 &$-$  &3.332  & 1.35 & $-$\\
	HSE06 Ref.~\cite{Kang2013}        &3.320  & 2.540   &$-$   & 1.33 & 1.14\\
	PBE Ref.~\cite{Horzum2013}      &3.321  & 2.528   &3.293  & 1.50 & $-$\\
	HSE Ref.~\cite{Kang2013}        &3.290   & 2.510  & $-$ &  1.52& 1.25\\
	GW Ref.~\cite{Kim2021} &$-$  &$-$  & $-$ & 2.08& $-$\\
	GVJ-2e Ref.~\cite{Gusakova} &$-$  &$-$  & $-$ & 2.03& $-$\\
	Experiment Refs~\cite{Lu2014,Shaw2014} &$-$  &$-$    & $-$  & \multicolumn{2}{c}{1.56}\\
	Experiment Ref.~\cite{Tonndorf2013}     & $-$  & $-$  & $-$ & \multicolumn{2}{c}{1.10}
\end{tabular}
\label{table:2D_parameters}
\end{ruledtabular}
\end{table*}

\subsection{$1H$-MoSe$_2$ monolayer with a Se vacancy}

\begin{table}[tb!]
\caption{Chemical potential $\mu_N(\mathrm{Se})$ related to the creation of one Se vacancy. $N\times N$, size of the supercell used in the calculations. 
$E_N(X)$, with $X=$MoSe$_2$, MoSe$_2$-vSe, and Se, corresponding cohesive energies provided by {\sc vasp}}
\label{table:finite_size_effects}
\begin{ruledtabular}
\begin{tabular}{ c c c c c }
	 \multicolumn{5}{c}{\textbf{revPBE-vdW}}\\	\hline	
	$N\times N$ & & $3\times3$ & $4\times4$ & $5\times5$	 \\\cline{3-5}			
	$E_N(\mathrm{MoSe2})/N^2$							& & $-$3.8599 &$-$3.8601 	&$-$3.8600  \\
	$E_N$(MoSe$_2$-vSe+Se)$/N^2$	& & $-$3.6720 &$-$3.7560 	&$-$3.7928 	\\
	$\mu_N(\mathrm{Se})$ 							&  & $-$5.07 & $-$5.02 	&$-$5.04 		\\
	\hline
	\\
	 	\multicolumn{5}{c}{\textbf{PBE-D3}}\\	\hline
	$N\times N$ & & $3\times3$ & $4\times4$ & $5\times5$ \\\cline{3-5}
	$E_N(\mathrm{MoSe2})/N^2$							& & $-$7.0051  	&$-$7.0052 	&$-$7.0052 \\
	$E_N$(MoSe$_2$-vSe+Se)$/N^2$  & &$-$7.0197  	&$-$7.0142 	& $-$7.0111  \\
	$\mu_N(\mathrm{Se})$ 							& & $-$5.82    &$-$5.78	&$-$5.76  
\end{tabular}
\end{ruledtabular}
\end{table}

Let us now consider the case in which Se vacancies are present in the MoSe$_2$ sheet (MoSe$_2$-vSe). Our aim is to describe the  limiting case in which the concentration of Se vacancies is so small that they can be considered as isolated. Therefore, as a first step we have to determine the size of the $(N\times N)$ supercell that assures a negligible interaction between the periodically repeated vacancies. 
To do so, we define a chemical potential $\mu_N(\mathrm{Se})$, related to the creation of an isolated Se vacancy, as
\begin{equation}
\mu_N(\mathrm{Se}) = E_N\mathrm{(MoSe_2)}-E_N\mathrm{(MoSe_2\mbox{-}vSe)} + E_N(Se),
\end{equation}
where $E_N(\mathrm{MoSe2})$, $E_N$(MoSe$_2$-vSe), and $E_N$(Se) are the {\sc vasp} total energies of MoSe$_2$, MoSe$_2$-vSe, and Se, respectively, that are calculated using the same $N\times N$ supercell. (The subscript $N$ refers to the size of the calculation cell). 

The $\mu_N(\mathrm{Se})$ values for different supercell sizes are summarized in Table~\ref{table:finite_size_effects}. In all cases the cell contains 23~{\AA} of vacuum along the surface normal. The linear extrapolation to $1/N\to0$ is $-$5.0 and $-$5.72~eV for revPBE-vdW and PBE-D3, respectively. The results in the table show that for the 4$\times$4 cell, finite-size effects in the chemical potential are below 1\%. Hence, this has been the supercell used in this work unless otherwise stated. 

Figure~\ref{fig:bands_vac} shows the MoSe$_2$-vSe band structure and DOS computed in the $4\times4$ supercell. Compared to the MoSe$_2$ band structure shown in the same figure, we clearly identify the in-gap states that appear above the Fermi level as a consequence of the Se vacancy, reducing in practice the band-gap energy. In the same figure, we show the projection of the total DOS into the $d$-states of Molybdenum and $p$-states of Selenium that dominate around the band-gap.  Note that the in-gap states are not completely degenerated in our case. In this respect, it is worth remarking the theoretical study of Ref.~\cite{Iberi2016} showing that the in-gap states break their degeneracy when the concentration of Se vacancies is finite. The splitting increases as the concentration of Se vacancies increases. Interestingly, for a vacancy concentration of 12\%, the authors found that the in-gap state is split into two bands all over the Brillouin zone except at the $\Gamma$-point. The splitting that we observe in our case is consistent to the one reported for our vacancy concentration of 3\%. 

"The presence of the Se vacancy induces a small shrink of the lattice around it. To evaluate that distortion, we calculate the ions displacement with respect to their (optimized) crystal positions in the pristine MoSe$_2$ monolayer. To avoid spurious finite size effects, we restrict ourselves to the atoms that lie at a distance from the vacancy position that is smaller than half the shortest dimension of the simulation cell. In the 4$\times$4 supercell this restriction corresponds to the first, second, and third Mo nearest neighbors and to the first and second Se neighbors. The average distortion that is induced at those five positions is $\sim$~0.13, 0.06, 0.01, 0.11, and 0.02~\AA, respectively."

\section{Atomic adsorption}
\label{app:at_ener}

\begin{table*}
\caption{Properties of the optimized configurations for adsorption of H, O, C, and N in MoSe$_2$ ($\mathrm{T_a}$ to $\mathrm{T_{ah}}$) and MoSe$_2$-vSe ($\mathrm{T_{vSe}}$): adsorption energy $E_a$, distance from the surface $z$ and total spin-magnetic moment $|\vec{M}|$. The sites that becomes unstable after the structural relaxation are marked with ``$-$''.}
\label{table:at_ener}
\begin{ruledtabular}
\begin{tabular}{l c c c c c c c c c c c c c c}

	&   &\multicolumn{6}{c}{\textbf{revPBE-vdW}} & & \multicolumn{6}{c}{\textbf{PBE-D3}}\\
	\cline{3-8}\cline{10-15}
atom	& & $\mathrm{T_a}$	& $\mathrm{T_b}$ & $\mathrm{T_h}$& $\mathrm{T_{ab}}$ & $\mathrm{T_{ah}}$& $\mathrm{T_{vSe}}$ & & $\mathrm{T_a}$	& $\mathrm{T_b}$ & $\mathrm{T_h}$& $\mathrm{T_{ab}}$ & $\mathrm{T_{ah}}$& $\mathrm{T_{vSe}}$ \\ 	\hline\\
	H & $E_a$(eV)	&  $-$0.388 & $-$0.115 &  $-$ & $-$0.446 & $-$0.689 &$-$2.715 & & $-$0.240 & $-$0.115 & $-$0.074 & $-$0.333 & $-$0.580 &$-$2.674\\
	& $z$(\AA)			 &  1.61 & 2.10&$-$& 1.52 & 1.30 & $-$0.84 & & 1.61& 1.85 & 2.65 &  1.45& 1.19 & $-$0.83 \\
	& $|\vec{M}|$($\mu_B$)	 	 & 1& 1& $-$ & 1& 1 & 1 & &1& 1& 1 & 1& 1 & 1   \\ \\
	\hline\\
	O & $E_a$(eV)	& -2.872	& $-$ & $-$ & $-$ & $-$ &  $-7.300$& &  $-3.230$ &  $-$0.313  & $-$ & $-$ & $-$ & $-$7.614 \\
	
	& $z$(\AA)			 & 1.60 & $-$ & $-$ & $-$ & $-$ &$-$0.67 & & 1.55 & 2.10 & & $-$ & $-$ & $-$0.62  \\
	& $|\vec{M}|$($\mu_B$)& 	0  & $-$ & $-$ & $-$ &  $-$  & 0 & & 0 & 2 & $-$ & $-$ & $-$ &  0  	 \\ \\
	\hline\\
	C & $E_a$(eV) & $-$1.617 & $-$2.178 & $-$& $-$& $-$ & $-$7.150 & & $-$1.990 & $-$2.756 & $-$& $-$& $-$ & $-$7.580	\\
	
	& $z$(\AA)	&  1.93 & 0.07 & $-$ & $-$ & $-$ &   $-$0.92 & & 1.82 & 0.03 &$-$ & $-$ & $-$ & $-$0.91   \\
	& $|\vec{M}|$($\mu_B$)	& 2& 2& $-$& $-$& $-$&0& & 2 & 2 & $-$ & $-$ &$-$  &0 \\ \\
	\hline\\
	N & $E_a$(eV)	& $-$0.712 &  $-$0.168 &  $-$0.172 &  $-$0.161 &  $-$0.163 & $-$6.259 & & $-$0.945 & $-$0.178 & $-$0.182 & $-$0.183 &  $-$0.180 & $-$6.512  \\
	
	& $z$(\AA)	& 1.75 & 2.77 & 2.75 & 2.88 & 2.88 & $-$0.77 & & 1.67	& 2.40	& 2.30	& 2.35 & 2.35 & $-$0.80    \\
	& $|\vec{M}|$($\mu_B$)& 1 & 3 & 3& 3 & 3 & 0 & & 1 & 3 & 3& 3 & 3 & 0 \\ \\
\end{tabular}
\end{ruledtabular}
\end{table*} 

In this appendix we analyze the adsorption on MoSe$_2$ of the different atoms that compose the diatomic molecules that we study in the main text: H$_2$, O$_2$, CO, and NO. The objective is to determine the most energetically favorable adsorption position among the five candidates that we propose, namely, $\mathrm{T_a}$, $\mathrm{T_b}$, $\mathrm{T_h}$, $\mathrm{T_{ab}}$, and $\mathrm{T_{ah}}$. Also following the main text, we study the atomic adsorption  at an isolated Se vacancy, $\mathrm{T_{vSe}}$.  Table~\ref{table:at_ener} summarizes the revPBE-vdW and PBE-D3 predictions for the adsorption energy $E_a$,  adsorbate distance from the surface $z$, and the total magnetic moment $|\vec{M}|(\mu_B)$. After relaxing the system, there are cases in which the adsorption position was unstable (the atom ends in another of the five sites), this is indicated in the table with the symbol ``-".

The most energetically favorable position for H is $\mathrm{T_{ah}}$ (in agreement with Ref.~\cite{Tsai2014}), with H adsorbed at a distance $z\sim$ 1.5 {\AA} from the surface and with $E_a$ = $-$0.689 ($-$0.580)~eV according to the revPBE-vdW (PBE-D3) calculation. It is worth remarking that while PBE-D3 predicts stability for all the five positions, the $\mathrm{T_{h}}$ position is unstable with revPBE-vdW. Adsorption at a Se vacancy is characterized by a much larger adsorption energy, $E_a^{\mathrm{MoSe_2}}(H)\sim$~$-$2.7~eV, with H located about 0.8~{\AA} below the MoSe$_2$-vSe surface. 

For Oxygen, both functionals predict that the most energetically favorable position is $\mathrm{T_{a}}$, with O  located at $z\sim$~1.6~{\AA} above the surface. The corresponding revPBE-vdW and PBE-D3 adsorption energies are $E_a=$~$-$2.872 and $-$3.230~eV, respectively. At variance with the revPBE-vdW results, 
PBE-D3 also predicts a weaker adsorption at the $\mathrm{T_b}$ position. In this case, the adsorbed O retains the spin-magnetic moment of 2 $\mu_B$.  
Adsorption at a Se vacancy with the O atom located at $Z\sim$ 0.65~{\AA} below the surface is much more favorable, being the revPBE-vdW (PBE-D3) adsorption energy $-$7.300 (-7.614)~eV.

In the case of Carbon, $\mathrm{T_b}$ with a revPBE-vdW (PBE-D3) calculated $E_a=-2.178$ ($-2.756$)~eV is the energetically preferred adsorption site, with C located at only 0.05~{\AA} from the surface. Furthermore, revPBE-vdW and PBE-D3 also predict a weaker adsorption at the $\mathrm{T_a}$ site. As in previous cases, adsorption at a Se vacancy is highly exothermic ($E_a=-7.2$~eV with revPBE-vdW and $E_a=-7.6$~eV with PBE-D3) and the C atom adsorbs below the surface ($z\sim-$0.9~{\AA}).

Nitrogen adsorption is energetically favorable in all the five positions that we have tested. However, the adsorption energy for $\mathrm{T_a}$ is more than four times larger than for any of the other positions. It is worth remarking that the total magnetic moment of the system is finite and located around the N atom in all the studied cases. However, only when N adsorbs on $\mathrm{T_a}$, the magnetic moment is reduced from the gas-phase value of 3 $\mu_B$ to $1 \mu_B$. As in all previous cases, adsorption at a Se vacancy is strongly favored, with $E_a=$~-6.256 (-6.512)~eV for revPBE-vdW (PBE-D3). In this situation, there is a Bader charge transfer to N of $\sim 1 e$ and no magnetic moment is appreciable around N. 

To conclude this appendix it is worth noticing that for O, C, and N, PBE-D3 predicts larger adsorption energies for the most energetically favorable configuration than revPBE-vdW. On the contrary, for H it occurs the opposite. In all cases the agreement in the structural configuration and magnetic state is good. 

Atomic adsorption on the pristine MoSe$_2$ surface was studied in Ref.~\cite{Ma2011} using PBE as the exchange-correlation functional. The reported adsorption energy and sites for H, O, C, and N are following this order: $-$0.021, $-$3.160, $-$2.599, and $-$0.714~eV at $\mathrm{T_{a}}$, $\mathrm{T_{a}}$, $\mathrm{T_{b}}$, and $\mathrm{T_{a}}$. For the last three cases, the PBE adsorption energies are within the PBE-D3 and revPBE-vdW values of Table~\ref{table:at_ener}. However, for the case of Hydrogen a much weaker adsorption energy is reported. We aatribute this discrepancy to the use of a different functional (PBE with no vdW correction) and, most probably, to the fact that only the $\mathrm{T_{a}}$, $\mathrm{T_{b}}$ and $\mathrm{T_{h}}$ positions were tested in their case.

\section{Spin-orbit coupling effect}
\label{sec:spin_orbit}

\begin{table*}[tb]
\caption{Effect of SOC in the PBE-D3 adsorption and dissociation energies.}
\label{table:energies_SOC}
\begin{ruledtabular}
\begin{tabular}{ r cccc c c c   c c }	
	&  \multicolumn{4}{c}{$E_a$(eV)}& &\multicolumn{4}{c}{$E_d$(eV)}\\
	\cline{2-5}\cline{4-10}
		 &\multicolumn{2}{c}{MoSe$_2$}	&  \multicolumn{2}{c}{MoSe$_2$-vSe} & 	& \multicolumn{2}{c}{MoSe$_2$}	&  \multicolumn{2}{c}{MoSe$_2$-vSe}\\
		 \cline{2-3}\cline{4-5}\cline{7-8}\cline{9-10}
	& No SOC & SOC& No SOC & SOC& & No SOC  & SOC	& No SOC		& SOC \\ 
	H$_2$ & $-$0.049  & $-$0.049  & $-$0.539   & $-$0.531 & & 3.378& 3.371 & 1.256 &1.259 \\
	O$_2$ & $-$0.172 & $-$0.086 &$-$2.266 &$-$2.260 & &   1.039 & 1.058 &$-$3.947 &$-$3.936\\
	CO & $-$0.145 & $-$0.126 &$-$1.317 &$-$1.317&  &   5.700 & 5.715 & 1.263 & 1.268\\
	NO & $-$0.204  & $-$0.205 &$-$2.778 &$-$2.772 & &  3.626 & 3.639&$-$2.043 &$-$2.048\\
\end{tabular}
\end{ruledtabular}
\end{table*} 

As discussed in Appendix~\ref{app:MoSe2_char}, SOC reduces the  MoSe$_2$ band-gap by about 100~meV (see Fig~\ref{fig:bands_vac}). The reason is the splitting that occurs at the conduction band minimum and valence band maximum  around the high symmetry $K$-point, as already confirmed by angle-resolved photo-electron spectroscopy~\cite{Boker2001} and transmission spectra~\cite{Beal1972} experiments. Our purpose in this appendix is to determine whether this reduction in the band gap may alter the adsorption and dissociation energies of the diatomic molecules we study in this work. For simplicity, the analysis is restricted to the PBE-D3 results and, unless otherwise stated, to the most energetically favored adsorption configurations. Each of the optimized PBE-D3 configurations are again relaxed with a PBE-D3 calculation that also includes SOC, using the same relaxation criteria as before. In Table~\ref{table:energies_SOC} we compare our results for the adsorption energy with and without SOC. For H$_2$, CO, and NO adsorbed on the pristine MoSe$_2$ surface, the differences in the adsorption energy calculated with and without SOC are small (i.e., $|E_a^{\mathrm{No\;SOC}} - E_a^{\mathrm{SOC}}|\le$~20~meV). However, in the case of O$_2$, the SOC calculation predicts a notable reduction in the adsorption energy ($E_a^{\mathrm{SOC}}=$-0.086~eV versus $E_a^{\mathrm{no\;SOC}}=$~-0.172~eV). On the other hand, when adsorption occurs at a Se vacancy, the SOC effect in $E_a$ is almost negligible in all cases.

Similarly, we systematically study the effect that SOC has in the dissociation energies of H$_2$, O$_2$, CO and NO. As in the previous case, the analysis is restricted to the most energetically favored dissociation configurations. Therefore, for dissociation of H$_2$, O$_2$, CO, and NO we consider following this order, $\mathrm{T_{ah,ah}}$(2H), $\mathrm{T_{a,a}}$(2O), $\mathrm{T_{b^\prime}(C)} + \mathrm{T_{ah}(O)}$, and $\mathrm{T_a(N)} + \mathrm{T_a(O)}$ in the case of MoSe$_2$ and $\mathrm{T_{vSe}}$(H) +$\mathrm{T_{ah}}$(H), $\mathrm{T_{vSe}}$(O) + $\mathrm{T_a}$(O), $\mathrm{T_{vSe}(C)} + \mathrm{T_a(O)}$  and $\mathrm{T_{vSe}(N)} + \mathrm{T_a(O)}$ in the case of MoSe$_2$-vSe. The effect in all cases is small being the largest one for the dissociation of O$_2$ on the pristine surface ($|E_d^{\mathrm{No\;SOC}} - E_d^{\mathrm{SOC}}|\sim$~20~meV). Although for simplicity we have not included it on the table, the difference $|E_d^{\mathrm{No\;SOC}} - E_d^{\mathrm{SOC}}|$ for the other dissociation positions that are commented in the main text is of the same order. All in all, this analysis suggests that the standard DFT calculations that neglect SOC can be safely used in our study.

\bibliography{MoSe2}

\end{document}